\RequirePackage{ifpdf}
\documentclass[12pt,a4paper,hyper]{JHEP3}
\usepackage{epsfig}
\usepackage{amsmath}
\usepackage{multirow}

\def\={\;  = \;}

\def\Slash#1{\rlap{\hbox{$\mskip 3 mu /$}}#1}      

\title{\center {Quantum entropy and exact 4D/5D connection}}

\preprint{}
\author{ Jo\~ao Gomes\\

\it DAMTP, Center for Mathematical Sciences\\
\it University of Cambridge, Wilberforce road\\
\it Cambridge, CB3 0WA, UK \\

{\rm Emails : jmg84 at cam.ac.uk\\
}}

\abstract{We consider the $AdS_{2}/CFT_{1}$ holographic correspondence near the horizon of rotating five-dimensional black holes preserving four supersymmetries in $\mathcal{N}=2$ supergravity.  The bulk partition function is given by a functional integral over string fields in $AdS_{2}$ and is related to the quantum entropy via the Sen's proposal. Under certain assumptions we use the idea of equivariant localization to non-rigid backgrounds and show that the path integral of off-shell supergravity on the near horizon background, which is a circle fibration over $AdS_2\times S^2$, reduces to a finite dimensional integral over $n_V+1$ parameters $C^A$, where $n_V$ is the number of vector multiplets of the theory while the $C^0$ mode corresponds to a normalizable fluctuation of the metric. The localization solutions, which rely only on off-shell supersymmetry, become after a field redefinition, the solutions found for localization of supergravity on $AdS_2\times S^2$.  We compute the renormalized action on the localization locus and show that, in the absence of higher derivative corrections, it agrees with the four dimensional counterpart computed on $AdS_2\times S^2$. These results together with possible one-loop contributions can be used to establish an exact connection between five and four dimensional quantum entropies. }


\newcommand{\bem}{\begin{pmatrix}}
\newcommand{\eem}{\end{pmatrix}}

\def\h{\eta}

\def\CN{{\cal N}}

\def\CN{{\cal N}}

\def\bea{\begin{eqnarray}}
\def\eea{\end{eqnarray}}
\def\be{\begin{equation}}
\def\ee{\end{equation}}
\def\ba{\begin{align}}
\def\ea{\end{align}}
\def\bse{\begin{subequations}}
\def\ese{\end{subequations}}
\def\1F1{{}_1\!F_1}
\def\2F0{{}_2\!F_0}

\def\h3{$\textrm{H}_3^+$}

\newcommand{\beq}{\begin{equation}}
\newcommand{\eeq}{\end{equation}}
\newcommand{\ber}{\begin{eqnarray}}
\newcommand{\eer}{\end{eqnarray}}

\def\be{\begin{eqnarray}}
\def\ee{\end{eqnarray}}

\def\CN{{\cal N}}

\font\manual=manfnt
\def\dbend{\lower3.5pt\hbox{\manual\char127}}

\def\bar{\overline}

\def\CN{{\cal N}}

\def\rt2{\sqrt{2}}
\def\irt2{{1\over\sqrt{2}}}

\font\cmss=cmss10
\font\cmsss=cmss10 at 7pt

\def\IL{\relax{\rm I\kern-.18em L}}
\def\IH{\relax{\rm I\kern-.18em H}}
\def\rlx{\relax\leavevmode}
\def\ZZ{\rlx\leavevmode\ifmmode\mathchoice{\hbox{\cmss Z\kern-.4em Z}}
 {\hbox{\cmss Z\kern-.4em Z}}{\lower.9pt\hbox{\cmsss Z\kern-.36em Z}}
 {\lower1.2pt\hbox{\cmsss Z\kern-.36em Z}}\else{\cmss Z\kern-.4em
 Z}\fi}

\newcommand{\ket}[1]{\left| #1 \right>} 
\newcommand{\bra}[1]{\left< #1 \right|} 
\newcommand{\braket}[2]{\left< #1 \vphantom{#2} \right|
 \left. #2 \vphantom{#1} \right>} 

\begin{document}

\section{Introduction}
In string theory or in any consistent quantum theory of gravity we should be able to describe a black hole as an ensemble of quantum states. The statistical entropy of the black hole or simply quantum entropy is given by the Boltzmann formula
\begin{equation}
S=\ln d(Q)
\end{equation}with $d(Q)$ the number of states with charge $Q$. In the thermodynamic limit or large charge regime the expression above is well approximated by the famous Bekenstein-Hawking area formula\footnote{In units where the Newton's constant $G=1$} \cite{Bekenstein:1973ur,Bekenstein:1983iq,Hawking:1974sw,Wald:1993nt,Iyer:1994ys,Jacobson:1993vj} 
\begin{equation}
S\simeq\frac{A}{4}
\end{equation} which then gives the semiclassical, leading contribution to the black hole's quantum entropy. Hawking's formula is in a sense very general and universal and therefore it does not tell much about the microscopic details of the theory. On the other hand, for extremal black holes, which have an $AdS_2$ near horizon geometry, finite charge corrections to the area formula can be used to test the holographic correspondence \cite{Maldacena:1997re} beyond the thermodynamic limit and infer details of the dual quantum theory. In this sense it is of great interest to compute finite charge corrections to the entropy and compare them for example with known contributions from BPS state counting. 

 Sen's proposal \cite{Sen:2008vm,Sen:2008yk} relates the quantum entropy of an extremal black hole to a path integral of string fields over $AdS_2$ with some Wilson line insertions at the boundary. Via the $AdS_2/CFT_1$ correspondence, it counts the number $d(Q)$ of ground states of the dual conformal quantum mechanics in a particular charge sector $Q$. The entropy is then given by the statistical formula $S=\ln d(Q)$. By putting the boundary of $AdS_2$ at finite  radius we generate an IR cuttoff \cite{Sen:2009vz} which can be used to extract relevant information via holographic renormalization. This definition then respects all the symmetries of the theory and reduces to the Wald formula in the corresponding limit of low curvatures or large horizon radius.  

The quantum entropy function  constitutes a powerful tool to compute finite charge corrections to the area law which can then be compared to microscopic calculations \footnote{By microscopics we mean the dual conformal quantum mechanics theory. In some contexts we count the BPS states by looking at the supersymmetric states of a 2d SCFT, like the D1-D5 low energy effective string.}. For many examples of supersymmetric black holes in both four dimensional $\CN=4,8$ string theories there are microscopic formulas for an indexed number\footnote{By indexed number of BPS states we mean a helicity trace index $B_n=\text{Tr}(-1)^hh^{n}$, where $h$ is the helicity quantum number and $n$ is the number of complex fermion zero modes} of BPS states \cite{Maldacena:1999bp,Dijkgraaf:1996it,Shih:2005qf,Dabholkar:2004yr,David:2006yn,Jatkar:2005bh,Sen:2008ta,Banerjee:2008pu,Dabholkar:2008zy}
valid in a large region of the charge configuration space. Using either the Cardy formula or an asymptotic expansion in the large charge limit \cite{Dabholkar:2010rm,Banerjee:2008ky}, the microscopic index agrees with the exponential of the Wald's entropy. Additional subleading corrections can then be compared with those obtained via the $AdS_2$ quantum entropy framework. For instance one-loop determinants of fluctuations of massless string fields over the attractor background  give logarithmic corrections to the area formula that are in perfect agreement with the microscopic answers \cite{Banerjee:2010qc,Sen:2011ba,Banerjee:2011jp} . 

Despite all this success, the techniques involved  are quite limited which makes the computation of further perturbative corrections an extremely difficult problem. However, for supersymmetric theories we hope to use localization to compute \emph{all of them exactly}. At least for rigid supersymmetric theories the principle is quite simple. We deform the original action by adding a $Q$-exact term of the form $tQV$, where $Q$ stands for some supersymmetry of the theory and the functional $V$ is chosen such that $Q^2V=0$. Then using the fact that both the action and the deformation are $Q$ invariant it can be shown that the path integral does not depend on $t$. So in the limit $t\rightarrow \infty$ the path integral collapses onto the saddle points of the deformation and the semiclassical approximation becomes exact. This explains the concept of localization. This technique  has been used extensively and with great success to compute exactly many observables in non-abelian gauge theories defined on a sphere \cite{Pestun:2007rz,Kapustin:2009kz} and recently many other cousins of these spaces. Recently the same technique was applied with great success to supergravity on $AdS_2\times S^2$ \cite{Dabholkar:2010uh} in the context of black hole entropy counting. A spectacular simplification was observed in that only a particular mode of the scalar fields was allowed to fluctuate, with the other fields fixed to their attractor values. We say that the path integral localized over a finite dimensional subspace of the phase configuration space. The renormalized action\footnote{This is the action of string fields on $AdS_2$ after removing IR cuttoff dependent terms.} has a very simple dependence on the prepotential of the theory  and is a function of $n_V+1$ parameters $C^I$ which have to be integrated. More recently in \cite{Dabholkar:2011ec} these results were applied in the case of four dimensional big black holes in toroidally compactified IIB string theory. The microscopic degeneracy, given as a fourier coefficient of a Jacobi form, can be rewritten in the Rademacher expansion and then compared with the gravity computation. The leading term of this expansion was reproduced exactly from these considerations. The non-perturbative corrections to this result, possibly coming from additional orbifolds, are more subleading, rendering the agreement between microscopics and macroscopics almost exact.

It would be interesting in the view of $AdS_2/CFT_1$ correspondence to test these ideas in other examples. The study of higher dimensional black holes in this context  is of particular interest for two main reasons. First, there is an interesting connection relating the microscopic partition functions of four and five dimensional black holes called $4d/5d$ lift \cite{Bena:2005ni,Bena:2004tk, Gaiotto:2005gf,Gaiotto:2005xt}. It would be very important to understand this connection from a bulk point of view at the quantum level. For instance the microscopic partition functions of four and five dimensional black holes in toroidally compactified string theory are the same \footnote{We are skipping issues related to hair contributions \cite{Banerjee:2009uk, Jatkar:2009yd}.}. Since the quantum entropies have to agree one expects the five dimensional theory to "reduce" to four dimensions exactly. Secondly, we want to understand how localization works in the presence of gravity, that is, in a non-rigid background. Since the four and five dimensional answers are related, it is expected that some mode of the five dimensional metric is left unfixed. As a matter of fact the near horizon geometry of a supersymmetric five dimensional black hole has the form of a circle fibered over $AdS_2\times S^2$ \cite{Gauntlett:1998fz} (which we denote as $AdS_2\times S^2\ltimes S^1$). The fiber, which carries angular momentum, gives rise to a $U(1)$ gauge field after dimensional reduction. Rigid supersymmetric localization is quite well understood. However localization in non-rigid backgrounds constitutes a new challenge and an interesting problem from a technical point of view. 

At the level of  two derivative supergravity action, the Bekenstein-Hawking entropy of the five-dimensional BMPV black hole \cite{Breckenridge:1996is} equals that of the four-dimensional supersymmetric black hole after identifying the five-dimensional angular momentum with four-dimensional electric charge. For $\mathcal{N}=8$ black holes in toroidally compactified string theory this equality should hold even at quantum level since the microscopic answers in $5d$\cite{Maldacena:1999bp} and $4d$ \cite{Sen:2008ta} are the same\footnote{The microscopic answers are the same except for a sign function $(-1)^J$, with $J$ the angular momentum.}. However in the case of $\mathcal{N}=4$ black holes the $4d/5d$ lift is non-trivial and the equality of quantum entropies is no longer true already at two derivative level\footnote{The logarithmic corrections computed using the two derivative supergravity action are different in four and five dimensional $\mathcal{N}=4$ theories.} \cite{Sen:2011cj} but always in agreement with the microscopic answers. In the view of the existing results for four-dimensional black holes in $\mathcal{N}=8$ string theory \cite{Dabholkar:2011ec} we give support for an exact bulk derivation of this property. We will find out that the renormalized action of both the five and four dimensional theories are the same. 

This work has therefore two fundamental purposes: to compute the partition function of supergravity on $AdS_2\times S^2\ltimes S^1$ using localization techniques and to establish a quantum version of the $4d/5d$ lift from a bulk perspective.  Instead of reducing the theory down to $AdS_2$, as in other perturbative computations \cite{Banerjee:2010qc,Sen:2011ba,Banerjee:2011jp}, we consider the five dimensional theory and apply localization to the off-shell $\mathcal{N}=2$ theory. Our work focus on the perturbative part of this computation, that is, in finding the saddle points of the localization action. We compute the renormalized action on the localization solutions in the case when higher derivative corrections are absent, which is appropriate for five dimensional $\mathcal{N}=8$ black holes. 

The use of localization in supergravity on $AdS_2\times S^2$, even though not fully understood, has remarkable results. As found in \cite{Dabholkar:2010uh}, the scalar in the compensating vector multiplet of four dimensional $\mathcal{N}=2$ off-shell supergravity is left unfixed by the localization equations. This means that from a five dimensional perspective we expect some mode of the metric to be left unfixed, namely the dilaton that measures the size of the fiber. In other words we need to consider localization on a non-rigid background. While it is straightforward to show localization in rigid supersymmetric theories this is not the case when the background itself is dynamical like in supergravity. The problem resides on the fact that it is very difficult to construct an exact deformation, if such an object exists, that is both gauge invariant and background independent. For rigid theories we usually pick a Killing spinor whose associated Killing vector generates a compact symmetry of the background. This is then enough to show exactness of the deformation. In general this choice of Killing spinor breaks the symmetries of the background and therefore it cannot be a diffeomorphic invariant deformation. We explain this in more detail later in section \S\ref{EqLocalz}. However, in supergravity this only makes sense in regions of the configuration space where we can use a partially fixed background. In these regions we can use a partially fixed Killing spinor that stills generates a compact symmetry. For instance we will show that by fixing the four dimensional metric to be $AdS_2\times S^2$ while leaving the fiber off-shell it is still possible to construct an exact deformation that can be used to localize the theory. Even though our understanding of localization in supergravity is only partial, our belief is that supergravity, at least on $AdS_2$ spaces, localizes, that is, there are just a finite number of modes that capture all quantum corrections. This is a strong statement but it is quite clear from the microscopic answers that something similar might be happening. 

In this sense we adopt a different approach in this work. We start from an ansatz for the metric and do the same for the Killing spinor. To be able to use localization we need a fermionic symmetry $\delta$ that generates a compact bosonic symmetry. More specifically we need
\begin{equation}\label{del^2}
\delta^2=\mathcal{L}_v+G
\end{equation}where $\mathcal{L}_v$ is the Lie derivative along a Killing vector $v$ and $G$ are gauge transformations. The operator $\delta$ has the name of twisted de Rham operator in differential geometry. As we will see later, in order for $\delta$ to act equivariantly, that is, in the sense of (\ref{del^2}), we need to impose certain conditions on the fields.

In order to use localization we need an off-shell realization of supersymmetry in five dimensions. A beautiful construction is given by the $\mathcal{N}=2$ five dimensional superconformal formalism developed recently in \cite{Bergshoeff:2004kh,Fujita:2001kv,Hanaki:2006pj,Bergshoeff:2001hc}. Although our interest is on BPS black holes in $\mathcal{N}=4,8$ theories, for which we have microscopic answers, we will use the $\mathcal{N}=2$ formalism where these black holes can be embedded.

The localization solutions, presented in section \S\ref{Sec Local in 5D}, look very complicated from a five dimensional perspective. There are a great number of fields both from the weyl multiplet and vector multiplets left unfixed by the localization equations. The hypermultiplet fields however remain fixed to their background values. This is only an assumption since we do not have an off-shell representation of their supersymmetric variations. After a field redefinition, these solutions are recognized to be the solutions found in \cite{Dabholkar:2010uh,Gupta:2012cy} for localization of supergravity on $AdS_2\times S^2$. They are parametrized by $n_V+1$ parameters $C^A$, with $n_V$ the number of vector multiplets, and label normalizable fluctuations of the four dimensional scalar fields.  These four dimensional scalar fields result from a combination of the five dimensional scalars $\sigma$ together with the fifth component $\chi$ of the gauge fields and a mode coming from the auxiliary antisymmetric field $T_{ab}$, in the Weyl multiplet, that we denote as $\alpha$, in the form
\begin{equation}
X^+=(\sigma+\chi)e^{\alpha},\,X^-=(\sigma-\chi)e^{-\alpha}.\nonumber
\end{equation}At asymptotic infinity $\alpha$ is related to the five dimensional angular momentum $J_{\psi}\propto \sinh(\alpha)$. The reduced euclidean theory  has $SO(1,1)$ R-symmetry which explains the use of paracomplex scalars $X_{\pm}$\footnote{Euclidean $\mathcal{N}=2$ supersymmetry in four dimensions has $SO(1,1)$ R-symmetry \cite{Cortes:2003zd}. The vector multiplet scalars are "charged" $\pm$ under this non-compact R-symmetry.}. The localization equations leave unfixed the "real" part of the paracomplex scalar fields which in hyperbolic coordinates\footnote{In hyperbolic coordinates $AdS_2$ metric is written as $ds^2=d\eta^2+\sinh(\eta)^2d\theta^2$.} have the following spacetime dependence
\begin{equation}
\text{Re}(X^+)=\frac{(X^++X^-)}{2}=(*)+\frac{C}{\cosh(\eta)}\nonumber
\end{equation}where $(*)$ denotes the on-shell value. On the other hand the dilaton  $\Phi$ which measures the size of the five dimensional circle combines also with $\alpha$ to give an additional paracomplex scalar
\begin{equation}
X_0^+=\Phi^{-1}e^{\alpha},\,X_0^-=\Phi^{-1}e^{-\alpha}
\end{equation}and gives an extra mode that has to be integrated out
\begin{equation}
\text{Im} (X_0)=\frac{X^+_0-X^-_0}{2}=\tanh(\alpha^*)+\frac{C_0}{\cosh(\eta)}. \nonumber
\end{equation}with $C_0$ an arbitrary constant. As in \cite{Dabholkar:2010uh} these modes can fluctuate if certain auxiliary fields are also allowed to fluctuate above their attractor values. The localization equations show that any dependence on the five dimensional coordinate drops out, rendering the $5d/4d$ reduction exact. However we need to be cautious about possible contributions from Kaluza Klein modes in the one-loop determinants but we do not consider this problem here. 

Note that the localization analysis uses off-shell supersymmetry and therefore it is independent of the particular details of the action. As explained later in section \S\ref{Bnd terms}, to guarantee that the supergravity action is invariant under the fermionic symmetry $\delta$ we need to add appropriate boundary terms, Wilson lines in this case. These boundary terms are important to guarantee a consistent variational principle. These Wilson lines are different from the electric Wilson lines used in \cite{Sen:2008vm,Sen:2008yk,Dabholkar:2010uh} as they do not carry explicit information about the five dimensional charges as we could have expected. Nonetheless, after some algebra, the renormalized action shows dependence on the five dimensional charges in a way that is consistent with the four dimensional results of \cite{Dabholkar:2010uh}. The final answer for the quantum degeneracy $d(Q)$, in the absence of higher derivative terms, is a finite dimensional integral over $n_V+1$ variables,
\begin{equation}
d(Q,J)=\int \prod_{I=0}^{n_V}d\phi^I \mathcal{M}(\phi)e^{S_{ren}(Q,J,\phi)}
\end{equation}where $\mathcal{M}$ denotes an effective measure on the space of $\phi^I$, $Q$ and $J$ are the charges and angular momentum respectively, and the renormalized action has the form
\begin{equation}
S_{ren}=\pi Q_I\phi^I+\pi J\phi^0-2\pi \displaystyle\frac{C_{IJK}\phi^I\phi^J\phi^K}{1-\phi_0^2}
\end{equation}where $C_{IJK}$ is a completely symmetric constant matrix. After a suitable analytic continuation of $\phi^0$ the renormalized action matches the four dimensional counterpart as expected from the $4d/5d$ lift. The measure $\mathcal{M}$ should in principle be computed from the one-loop determinants. However since the localization equations are valid only in a region of the full configuration space we do not have access to all fluctuations orthogonal to the localization locus but only a subset.

The paper is organized as follows. In section \S\ref{QEF} we review the quantum entropy function formalism. In section \S\ref{EqLocalz} we explain the technique of equivariant localization starting from finite dimensional integrals and then introducing the case of infinite dimensional integrals. In section \S\ref{Sec: superconf} we review the five dimensional superconformal formalism. We introduce the supersymmetric variations of the various supermultiplets, the lagrangian and also the full BPS attractor equations. Finally in section \S\ref{Sec Local in 5D} we do localization of the supergravity theory and compute the renormalized action on the localization locus.

\section{Quantum entropy function and $AdS_2/CFT_1$ correspondence}\label{QEF}

The quantum entropy function \cite{Sen:2008yk,Sen:2008vm,Sen:2011cn,Sen:2009vz},  based on the $AdS_2/CFT_1$ correspondence, is a proposal for the quantum entropy of an extremal black hole. By quantum entropy we mean a generalization of the Wald's formula that  captures the entropy of a black hole described as an ensemble of quantum states. At least for BPS black holes for which there are precise microscopic answers it is believed that such a formula might exist. We have to stress the fact that notwithstanding the microscopic answer being an index, it has been argued in \cite{Sen:2009vz,Dabholkar:2010rm} that for black holes that preserve at least four supercharges index equals degeneracy for the near horizon degrees of freedom. 

 The quantum degeneracy $d(q)$, where $q$ labels the charges of the black hole, counts the number of ground states of the conformal quantum mechanics and is related, via $AdS/CFT$ correspondence, to the partition function of supergravity on $AdS_2$ with Wilson lines inserted at the boundary:
\begin{equation}\label{ExpctWilsonLine}
d(q)=\langle e^{-iq\oint A}\rangle_{AdS_2}
\end{equation}where the geometry has euclidean signature\footnote{As usual we perform a Wick rotation $t\rightarrow -i\theta$. In other instances we learned to take $t\rightarrow i\theta$ such that the path integrand becomes $e^{-S}$ with $S$ positive providing a convergent integral. Here however the euclidean action is already divergent due to the infinite volume of $AdS_2$ and it is the renormalized action that provides the correct damping exponential. In short, $\text{Ren}\,e^{S}=e^{-S_{ren}}$ }.

The Wilson line insertions can be understood from two different but equivalent ways. From an holographic point view, the electric part of the gauge fields carries a non-normalizable component at asymptotic infinity, that is, in coordinates where the boundary is at $r\rightarrow \infty$ the gauge field goes as $A\sim e r$, and therefore via the usual bulk/boundary correspondence dictionary \cite{Witten:1998qj} these modes have to be fixed while the normalizable component, the chemical potentials, have to be integrated out. This is in contrast with higher dimensional examples like in $AdS_4$. The microcanonical ensemble is natural from this point of view. But we can also see these Wilson lines as a requirement of a consistent formulation of the path integral. Without the Wilson lines the equations of motion for the gauge fields are not obeyed at the boundary because they carry a non-normalizable component \footnote{In other words the boundary terms that arise from varying the action do not vanish at asymptotic infinity}. Much like the Gibbons-Hawking terms, the path integral on $AdS_2$ requires appropriate boundary terms, Wilson lines in this case, that restore the validity of the equations of motion throughout all the space. We develop this idea further in section \S\ref{Bnd terms}.

This formalism surpasses in many ways other attempts to compute quantum corrections to the entropy. The success comes essentially from two basic facts. Firstly, there is a natural UV cutoff, the string scale $l_s$. Secondly, it introduces via holographic renormalization an IR cutoff which is essential for extracting relevant information even at the classical level. Besides, this formalism respects all the symmetries of string theory and reduces to Wald's formalism in the limit of low curvatures or large horizon. To see this consider the following simple example. The relevant near horizon data of an extremal black hole is given by the metric
\begin{equation}\label{attractor backgr 1}
ds^2=v(r^2-1)d\theta^2+v\frac{dr^2}{r^2-1}
\end{equation} with conformal boundary at $r\rightarrow\infty$, and gauge fields and scalar fields
\begin{equation}\label{attractor backgr 2}
A_{\mu}dx^{\mu}=-ie(r-1)d\theta,\; \Phi=constant,
\end{equation}respectively. The leading contribution to (\ref{ExpctWilsonLine}) comes from evaluating the action on the configuration described above. Since $AdS_2$ has infinite volume we introduce a cutoff at $r=r_0$ and discard terms that are linearly divergent, that is,
\begin{equation}\label{RenEntropyFnct}
\langle e^{-iq\oint A}\rangle_{AdS_2}\simeq \text{Ren}\lbrace e^{-2\pi q e(r_0-1)+(r_0-1)2\pi \mathcal{L}(v,e,\Phi)}\rbrace=e^{2\pi q e-2\pi \mathcal{L}(v,e,\Phi)}
\end{equation}where $\text{Ren}$ denotes renormalization by appropriate boundary counter terms that remove the $r_0$ dependence. The most RHS expression is just the exponential of the Wald's entropy. If we want to compute quantum contributions we look at normalizable fluctuations order by order in perturbation theory. This is essentially the work done in \cite{Banerjee:2010qc,Sen:2011ba,Banerjee:2011jp,Sen:2011cj}. The authors consider the reduced theory on $AdS_2$ and look at normalizable fluctuations of the background (\ref{attractor backgr 1}) and (\ref{attractor backgr 2}).
They compute, using the heat kernel method, one-loop determinants in the two derivative supergravity action. These terms give corrections of order $\ln(A)$ to the entropy, where $A$ is the horizon area in appropriate units.

When performing localization we will consider the path integral defined on the five-dimensional space $AdS_2\times S^2\ltimes S^1$ instead of reducing all the fields down to $AdS_2$. This method is obviously favorable for a number of reasons. However in the context of localization it requires the addition of appropriate boundary terms. Some of them arise by demanding that the equations of motion be obeyed also at the boundary, as explained before, others are necessary to restore gauge invariance. As we will explain later in the section \S \ref{EqLocalz} these terms are necessary for invariance of the action under supersymmetry, an essential ingredient for using localization. This condition will allow us to define in five dimensions an entropy function \textit{\`{a} la} Sen \cite{Sen:2005wa}.  Current available attempts \cite{Cardoso:2007rg} circumvent this problem by reducing to four dimensions which is clearly unsatisfactory in the view of our main goal.

\section{Localization Principle}\label{EqLocalz}

For illustrative purposes consider the  example of a finite dimensional integral $I(t)$ over a compact manifold $\mathcal{M}$,
\begin{equation}
I(t)=\int_{\mathcal{M}} e^{itf(m)}dm.
\end{equation} If $f(m)$ has a finite number of non-degenerate critical points $p\in\{f'(m)=0\} $, a saddle point approximation gives an asymptotic expansion in $t$
\begin{equation}
I(t)\sim \sum_{p\in \{ f'(m)=0\}}e^{itf(p)}\sum_{l\geq 0}a_l(p) t^{k-l}
\end{equation}where the coefficients $a_l(p)$ can be computed in terms of $f(m)$. 

For a certain class of integrals an extraordinary simplification occurs. If $\mathcal{M}$ is a symplectic manifold and $f(m)$ is the hamiltonian of an $S^1$ action on $\mathcal{M}$ then the "higher loop" corrections, that is, the $l>1$ terms vanish and the saddle point approximation becomes exact. This is the simplified version of the Duistermaat-Heckman theorem \cite{Duistermaat:1982xu}. In such a case, the integral localizes exactly over the critical points of $f(m)$ which are also the fixed points of the $S^1$ action on $\mathcal{M}$
\begin{equation}
I(t)=\sum_{p\in \{V=0\}}\frac{e^{itf(p)}}{\sqrt{\text{det}f''(p)}}t^{-\frac{n}{2}}
\end{equation}where $2n$ is the dimension of $\mathcal{M}$ and $V$ is the vector associated with the $S^1$ action. The fact that the integral only depends on the neighborhood data of the fixed points is commonly referred as localization. 

To better understand the mechanics of localization we need to study equivariant cohomology. This is particularly well understood for finite dimensional integrals. A good mathematical reference is \cite{Vergne} while \cite{Szabo:2000fs} is more convenient for a physicist point of view.  

Without entering in too many mathematical details we will explain briefly localization of finite dimensional integrals. The idea behind localization is that is possible to define on a manifold $\mathcal{M}$ an operator $\mathcal{D}$ which has the property that it squares to an isometry of the space. In other words
\begin{equation}
\mathcal{D}^2=\mathcal{L}_V
\end{equation}where $\mathcal{L}_V$ is the Lie derivative. On the space of forms the operator $\mathcal{D}$, also called twisted de Rham differential, has the form
\begin{equation}
\mathcal{D}=d+i_V
\end{equation}where $d$ is the de Rham differential and $i_V$ is the contraction operator by the vector $V$. Since the vector generates an isometry of the manifold we have $\mathcal{L}_Vg_{mn}=0$, that is, $V$ is a Killing vector. This operator then allows to define a cohomology on the space of forms which are left invariant under the isometry generated by $V$, that is, on the space of forms $\alpha$ for which $\mathcal{L}_V\alpha=0$. These forms are also called equivariant forms.
 
 It turns out that the integral over $\mathcal{M}$ of a closed form $\alpha$, that is, $\mathcal{D}_V\alpha=0$, localizes on the fixed points of action of $V$.  
To show localization we consider the auxiliary integral parametrized by $t$
\begin{equation}\label{Z(t)}
Z(t)=\int_{\mathcal{M}}\alpha e^{-t \mathcal{D}_V\beta}
\end{equation}with $\beta$ an equivariant differential form. Since both $\alpha$ and $\mathcal{D}_V\beta$ are closed under $\mathcal{D}_V$ we can show by integration by parts that
\begin{equation}\label{t independence}
\frac{dZ(t)}{dt}=0, \forall t
\end{equation}or in other words $\mathcal{D}_V\beta$ is an exact deformation. A clever choice for $\beta$ is to take
\begin{equation}
\beta=g_{\mu\nu}V^{\nu}(x)dx^{\mu},
\end{equation}with $g_{\mu\nu}$ a metric on the manifold $\mathcal{M}$. The property that $\mathcal{L}_Vg_{\mu\nu}=0$ ensures that $\beta$ is an equivariant form. That is a clever choice because the "deformation" $\mathcal{D}_V\beta$ has a term which is positive everywhere on $\mathcal{M}$
\begin{equation}
\mathcal{D}_V{\beta}=d\beta+V^2.
\end{equation}This can be used to show that in the limit $t\rightarrow \infty$ the integral collapses onto the fixed points of $V$, rendering the saddle point approximation exact.	This is the equivariant localization principle. 

The same idea can be applied to infinite dimensional integrals. The idea is to extend the properties of the operator $\mathcal{D}$ to the space of fields. Since it mixes forms of even and odd degrees it behaves much like a supercharge in supersymmetric field theories that sends bosonic to fermionic fields and vice-versa.  The twisted de Rham operator becomes a functional and can be identified with the action of a real supersymmetric transformation while the  analog of a closed equivariant form is given by a supersymmetric functional. By the same token we can deform the integral by an exact equivariant functional and show localization of the theory. 

To make things simple consider the case of one-dimensional $N=1/2$ supersymmetric quantum mechanics on a circle with period $T$ \cite{Szabo:2000fs}.

There exists a supersymmetry $\mathcal{S}$ that takes a boson to a fermion and vice-versa
\begin{equation}
\mathcal{S}X(\tau)=\Psi(\tau),\;\mathcal{S}\Psi(\tau)=\dot{X}
\end{equation}with $\tau$ the coordinate on the circle. These transformations can be used to define the functional equivariant operator
\begin{equation}
\mathcal{D}_{\dot{x}}=\mathcal{D}+\mathcal{I}
\end{equation}with
\begin{equation}
\mathcal{D}=\oint d\tau\Psi\frac{\delta}{\delta X(\tau)},\,\text{and }\mathcal{I}=\oint d\tau \dot{X}\frac{\delta}{\delta \Psi(\tau)}
\end{equation}It is an easy exercise to show that this operator squares to translations as expected from the supersymmetry algebra $Q^2=H$, with $H$, the hamiltonian,
\begin{equation}
\mathcal{D}_{\dot{x}}^2=\oint d\tau\left[\dot{X}\frac{\delta}{\delta X}+\dot{\Psi}\frac{\delta}{\delta \Psi}\right]=\oint d\tau\frac{d}{d\tau}.
\end{equation}The space of equivariant functionals is determined by functionals $W[X,\Psi]$ that vanish under the action of $\mathcal{D}_{\dot{x}}^2$. This immediately gives the condition
\begin{equation}\label{EquivPrinciple}
\mathcal{D}_{\dot{x}}^2W[X,\Psi]=\oint d\tau \frac{d}{d\tau}W[X,\Psi]=W\left[X(T),\Psi(T)\right]-W\left[X(0),\Psi(0)\right]=0.
\end{equation}Since this should be valid for any $X$ and $\Psi$ the condition is satisfied if we impose periodic conditions on both the scalars $X$ and fermions $\Psi$. In other words the \textit{space of equivariant functionals is the space of functionals with both $X$ and $\Psi$ fields periodically indentified on the circle}. The localization principle follows analogously. We deform the original integral by adding an exact deformation to the action
\begin{equation}\label{eqLoc:exact deformation}
S[X,\Psi]\rightarrow S[X,\Psi]-t\mathcal{D}_{\dot{x}}\oint d\tau \delta\Psi(\tau)\Psi(\tau).
\end{equation}Using the fact that the  deformed integrand is equivariantly closed we can show as in (\ref{t independence}) that the integral
\begin{equation}
I=\int DXD\Psi e^{S[X,\Psi]-t\mathcal{D}_{\dot{x}}\oint d\tau \delta\Psi(\tau)\Psi(\tau)}
\end{equation}does not depend on the parameter $t$ and consequently the limit $t\rightarrow \infty$ can be used to prove localization of the theory on the space of configurations for which
\begin{equation}
\delta \Psi=0\Leftrightarrow \dot{X}(\tau)=0.
\end{equation}That is, the theory localizes on constant fields $X$. Further corrections, which include the contribution from the Kaluza-Klein modes, are one-loop exact.

Without much effort the same idea can be applied  to higher dimensional supersymmetric theories. In general, localization in rigid supersymmetric gauge theories is quite straightforward as long as there is a fermionic symmetry that squares to a compact Killing symmetry of the background. More generally there is an odd symmetry $\delta$ with the property that
\begin{equation}\label{delta squared}
\delta^2=\mathcal{L}_v+G_a
\end{equation}where $v^{\mu}$ is a Killing vector and $G_a$ denotes a gauge transformation with parameter $a$. With a set of fields that respect this algebra we can easily construct an exact deformation of the physical action. For this reason any deformation of the form $\delta W$ with $W$ gauge invariant and $\partial_vW=0$, will be an exact deformation if the fields respect periodic boundary conditions along the compact $v$ direction. In other words
\begin{equation}\label{EquivCond}
\delta^2W[X,\Psi]=\int\mathcal{L}_v W(X,\Psi)=\int dv \partial_v W=0
\end{equation}

Pestun in his seminal work \cite{Pestun:2007rz} gives a beautiful application of this formalism in the computation of Wilson loops in $\mathcal{N}=4,2$ SYM defined on $S^4$. In this case he uses a fermionic symmetry which is a combination of a conventional Q-supersymmetry and a special S-supersymmetry. This fermionic symmetry squares to an antiself-dual rotation of the sphere plus $R$-symmetry and gauge transformations. 

For non-rigid supersymmetric theories it is not known if the same idea can be applied. We do not know how to construct, if it exists, an exact deformation that is both gauge invariant and background independent. In general it is difficult to find an odd symmetry that satisfies the condition (\ref{delta squared}). The case is even worst when there is gravity. However in a certain region of the phase configuration space it is possible to realize linearly such an algebra. For instance the authors in \cite{Dabholkar:2010uh} claim to have computed exactly the path integral of $\mathcal{N}=2$ supergravity on $AdS_2\times S^2$. The results are quite astonishing. Assuming that the background remains fixed they localize the gauge theory sector and find that for each vector multiplet a normalizable fluctuation of the scalars is allowed if the corresponding auxiliary scalar also fluctuates. They have found that the theory localizes on the set of fluctuations of the form
\begin{equation}
X=X^*+\frac{C}{r},\,\bar{X}=\bar{X}^*+\frac{C}{r}\,\text{with }K=\frac{C}{r^2}
\end{equation}with $X$ a scalar and $K$ the auxiliary scalar field, in the coordinates (\ref{attractor backgr 1}). Integration over the constants $C$ yields a finite dimensional integral which agrees with the microscopic predictions for $1/8$ BPS black holes in $\mathcal{N}=8$ string theory \cite{Dabholkar:2011ec}. 

Since in general the susy transformations of supergravity do not respect equivariant properties, the strategy that we pursue here is to find in which region of configuration space those properties are realized. This brings additional constraints on the fields. On this restricted subspace we can deform the path integral and show localization.  We believe that in the full quantum gauge fixed theory such a restriction would follow naturally.

\section{$5D$ superconformal gravity and near horizon analysis}\label{Sec: superconf}

In this section we introduce the $\mathcal{N}=2$ off-shell superconformal formalism for five dimensional supergravity. We present the various multiplets and respective supersymmetric transformations. We introduce the lagrangian with supersymmetric higher derivative corrections and present the BPS attractor equations for the $AdS_2\times S^2\ltimes S^1$ near horizon geometry of the BMPV black hole.

\subsection{Superconformal formalism}
The superconformal calculus was originally constructed for $\mathcal{N}=2$ supergravity in four dimensions \cite{deWit:1984px,deWit:1980tn,deWit:1979ug} but only recently a formulation in five dimensions was developed \cite{Bergshoeff:2004kh,Fujita:2001kv,Bergshoeff:2001hc,Hanaki:2006pj}. The idea is to construct a supersymmetric theory for the five dimensional conformal group by gauging the global generators and then imposing appropriate gauge fixing conditions. This is similar to the example of a scalar conformally coupled to the Einstein-Hilbert term. By gauge fixing the scalar to a constant we recover Poincar\'{e} gravity. 
One major distinction between four and five dimensional formulations is that while the first has $SU(2)\times U(1)$ R-symmetry, the five dimensional theory only has $SU(2)$ R-symmetry. This means, for instance, that the scalars in the vector multiplets are real.

 In the following we give a summary of the content of the various supermultiplets, namely the Weyl multiplet, the vector multiplet, the linear multiplet and the hypermultiplet, and respective supertransformation rules. We follow closely the paper \cite{deWit:2009de} where more details can be found. 
 
 We denote coordinate indices by greek letters $\mu,\nu,\ldots$, tangent space indices by roman letters $a,b,\ldots$ and R-symmetry $SU(2)$ indices by $i,j$.
 
\begin{description}
\item[Weyl multiplet:] the independent fields consist of the funfbein $e^a_{\mu}$, the gravitino field $\psi_{\mu}^i$, the dilatational gauge field $b_{\mu}$, the R-symmetry gauge fields $V_{\mu j}^i$ (anti-hermitian traceless matrix in the $SU(2)$ indices $i,j$), a real tensor field $T_{ab}$, a scalar $D$ and a spinor field $\chi^i$. Both $V_{\mu j}^i$, $T_{ab}$, $D$ and $\chi^i$ are auxiliary fields. For the problem we want to solve we set $b_{\mu}=0$ and gauge the special conformal transformations $K_a$ parameters $\Lambda_{K}^a$ to zero. The conventional $Q$ and special $S$ supersymmetry transformations, parametrized respectively by the spinors $\xi^i$ and $\eta^i$, are as follows:
\begin{eqnarray}\label{Weylvar}
\delta e^a_{\mu}&=&\frac{1}{2}\bar{\xi}_i\gamma^a \psi_{\mu}^i\nonumber\\
\delta \psi^i_{\mu}&=&D_{\mu}\xi^i+\frac{1}{2}V_{\mu j}^{\;\;\;i}\xi^j+\frac{i}{4}T_{ab}(3\gamma^{ab}\gamma_{\mu}-\gamma_{\mu}\gamma^{ab})\xi^i-i\frac{1}{2}\gamma_{\mu}\eta^i\nonumber \\
\delta V_{\mu i}{}^j &=& 3 i
  \bar\xi_{i} \phi_{\mu}{}^{j}
  -8\bar\xi_{i}\gamma_\mu\chi^{j} -3 {i}
  \bar\eta_{i}\psi_\mu{}^{j} + 
  \delta^i{}_j\,[-\frac{3}{2}{i}\bar\xi_{k}\phi_{\mu}{}^{k}
  +4\bar\epsilon_{k}\gamma_\mu\chi^{k}+\frac{3}{2}{i}
  \bar\eta_{k}\psi_\mu^{k}] \,, \nonumber \\  
   \delta T_{ab} &=&  \frac{2}{3} {i} \bar\xi_i \gamma_{ab}
  \chi^i -\frac{1}{8} {i} \bar\xi_i R_{ab}{}^i(Q)\,,
  \nonumber\\  
  \delta \chi^i &=&  
  \frac{ 1}{4} \xi^i D +\frac{1}{128} 
  R_{\mu\nu j}{}^{i}(V) \gamma^{\mu\nu} \epsilon^j 
  + \frac{3}{128} {i}(3\, \gamma^{ab} \Slash{D}
  +\Slash{D}\gamma^{ab})T_{ab} \, \xi^i \nonumber\\
  &&{}
  -\frac{3}{32} T_{ab}T_{cd}\gamma^{abcd}\xi^i                                   
  +\frac{3}{16} T_{ab}\gamma^{ab} \eta^i  \,, \nonumber\\
   \delta D &=&{}
  \bar\xi_i \Slash{D} \chi^i - {i}
  \bar\xi_i  T_{ab}\gamma^{ab} \chi^i - {i}
  \bar\eta_i\chi^i \,. 
\end{eqnarray} The derivatives $D_{\mu}$ are covariant derivatives.
\item[Vector multiplet:] the vector multiplet consists of a real scalar $\sigma$, a gauge field $W_{\mu}$, a triplet of auxiliary fields $Y^{ij}$ and a fermion field $\Omega^i$. The superconformal transformations are as follows:
\begin{eqnarray}\label{VectorSusyVar}
\delta \sigma &=&{}
  \frac{1}{2}{i}\bar{\xi}_i\Omega^i \,,
  \nonumber \\ 
  \delta\Omega^i &=&{}
  - \frac{1}{4} (\hat{F}_{ab}- 4\,\sigma T_{ab}) \gamma^{ab} \xi^i
  -\frac{1}{2} {i} \Slash{D} \sigma\xi^i -\varepsilon_{jk}\,
  Y^{ij} \xi^k 
  + \frac{1}{2}\sigma\,\eta^i \,,   \nonumber\\ 
  \delta W_\mu&=&{}
  \frac{1}{2} \bar{\xi}_i\gamma_\mu\Omega^i -\frac{1}{2} {i}
  \sigma \,\bar\xi_i \psi^i_\mu \,, \nonumber\\ 
  \delta Y^{ij}&=&{}  
  \frac{1}{2} \varepsilon^{k(i}\, \bar{\xi}_k \Slash{D} \Omega^{j)} 
  + {i}\varepsilon^{k(i}\, \bar\xi_k (-\frac{1}{4}
  T_{ab}\gamma^{ab}\Omega^{j)}+ 4 \sigma \chi^{j)})
  -\frac{1}{2} {i}  \varepsilon^{k(i}\, \bar{\eta}_k
  \Omega^{j)} \,.  
\end{eqnarray}with $Y_{ij}=\varepsilon_{ik}\varepsilon_{jl}
Y^{kl}$, and the supercovariant field strength is defined as, 
\begin{equation}
 \hat F_{\mu\nu}= \partial_\mu W_\nu - \partial_\nu W_\mu -
\bar\Omega_i\gamma_{[\mu} \psi_{\nu]}{}^i +\frac{1}{2}
\mathrm{i}\sigma\,\bar\psi_{[\mu i} \psi_{\nu]}{}^i \,.
\end{equation}

\item[Linear multiplet:] though they do not play any relevant role in our work we decided to include the supersymmetric transformations of the linear multiplet for congruence of the exposition. The linear multiplet consists of a triplet of scalars $L^{ij}$, a divergence-free vector $\hat{E}^a$, an auxiliary scalar $N$ and a fermion field $\varphi^i$. The superconformal transformations are as follows:
\begin{eqnarray}
 \delta L^{ij}&=&{}
  -\mathrm{i} \,\varepsilon^{k(i}\,\bar\xi_k \varphi^{j)}
  \,,  \nonumber\\
  \delta \varphi^{i}&=&{}
  -\frac{1}{2} {i}\, \varepsilon_{jk}\Slash{D} L^{ij}\xi^k+ \frac{1}{2}
  (N-{i}\hat{\Slash{E}})\xi^i 
  + 3\varepsilon_{jk} L^{ij} \eta^k  \, ,\nonumber\\ 
  \delta \hat E_a &=&{}
  -\frac{1}{2} {i}\, \bar{\xi_{i}}\gamma_{ab} D^b \varphi^{i}
  +\frac{1}{8} 
  \bar{\xi_{i}}(3\gamma_a\gamma^{bc} + \gamma^{bc}\gamma_a)
  \varphi^{i}T_{bc}  -2 
  \bar\eta_{i}\gamma_a\varphi^{i} \, ,\nonumber\\ 
  \delta N &=&{}
  \frac{1}{2}\bar{\xi_i} \Slash{D} \varphi^i + \frac{3}{4} {i}
  \bar\xi_i \gamma^{ab}\varphi^iT_{ab} -4 {i} \,
  \varepsilon_{jk}\,\bar{\xi_i}\chi^k L^{ij}
  +\frac{3}{2}  {i}\bar\eta_i\varphi^i  \,. 
\end{eqnarray} The divergence free condition of $\hat{E}^{\mu}$ can be easily solved by considering the three-rank antisymmetric tensor $E_{\mu\nu\rho}$ via the equation $\hat{E}=*dE$.

\item[Hypermultiplet:] hypermultiplets are usually associated with target spaces of dimension $4r$ that are hyperkahler cones. The superconformal transformations are written in terms of local sections $A_i^{\alpha}$ of an $Sp(r)\times Sp(1)$ bundle as follows
\begin{eqnarray} 
  \delta A_i{}^\alpha&=& {i}\,\bar\xi_i\zeta^\alpha\,,
  \nonumber\\ 
  \delta\zeta^\alpha &=& -\frac{1}{2} {i}\Slash{D}
  A_i{}^\alpha\xi^i 
  + \frac{3}{2} A_i{}^\alpha\eta^i \,.
\end{eqnarray}The covariant derivative contains the $Sp(r)$ connection $\Gamma^{\alpha}_{a\beta}$ associated with rotations of the fermions. Moreover the sections $A^{\alpha}_{i}$ are pseudo-real in the sense that they obey the constraint $(A_j^{\alpha})^*\equiv A^j_{\alpha}=A_i^{\beta}\varepsilon^{ij}\Omega_{\beta \alpha} $, where $\Omega_{\alpha\beta}$ is a covariantly constant skew-symmetric tensor with its complex conjugate satisfying $\Omega_{\alpha \beta}\Omega^{\beta \gamma}=\delta^{\gamma}_{\alpha}$. The information on the target space metric is contained in the hyperkahler potential
\begin{equation}
\varepsilon_{ij}\chi=\Omega_{\alpha\beta}A^{\alpha}_iA^{\beta}_j
\end{equation}
Note that the hypermultiplets do not exist as an off-shell supermultiplet. The superconformal transformations close only up to fermionic equations of motion.
\end{description}

\subsection{The Lagrangian}
We present the bosonic part of the Lagrangian.

The lagrangian is essentially the sum of three parcels, that is,
\begin{equation}
\mathcal{L}=\mathcal{L}_{VVV}+\mathcal{L}_H+\mathcal{L}_{VWW}
\end{equation} The first term $\mathcal{L}_{VVV}$ is cubic in the vector multiplet fields,
\begin{eqnarray}
 8\pi^2\mathcal{L}_{VVV} &=&{}
  3\,C_{IJK} \sigma^I\Big[\frac{1}{2} {D}_\mu\sigma^J
  \,{D}^{\mu}\sigma^K 
  + \frac{1}{4} F_{\mu\nu}{}^J F^{\mu\nu K} - Y_{ij}{}^J  Y^{ijK }
  -3\,\sigma^J F_{\mu\nu}{}^K  T^{\mu\nu} \Big] \nonumber \\
  &&{}
   - \frac{1}{8} {i}C_{IJK}\,e^{-1}
  \varepsilon^{\mu\nu\rho\sigma\tau} W_\mu{}^I 
  F_{\nu\rho}{}^J F_{\sigma\tau}{}^K 
  - C(\sigma) \Big[-\frac{1}{8} \mathcal{R} - 4\,D - \frac{39}{2}
  T^2\Big]\,, 
\end{eqnarray}where $C_{IJK}$, symmetric in all its indices, are constants that encode the different couplings of the fields. The function $C(\sigma)$ is the contraction $C(\sigma)=C_{IJK}\sigma^I\sigma^J\sigma^K$.

The term $\mathcal{L_H}$ encodes the lagrangian for the hypermultiplets
\begin{equation}\label{Lagr Hyper}
8\pi^2\mathcal{L}_H=  -\frac{1}{2}
  \Omega_{\alpha\beta}\, \varepsilon^{ij} 
  { D}_\mu A_i{}^\alpha\, { D}^\mu
  A_j{}^{\beta}+ \chi\Big[-\frac{3}{16}\mathcal{R} 
  + 2\, D  + \frac{3}{4} T^2 \Big]\,,
\end{equation}while $\mathcal{L}_{VWW}$ contains higher derivative corrections with couplings between vector and weyl multiplets fields
\begin{eqnarray}
  8\pi^2\mathcal{L}_{VWW} &=&{}
  +\frac{1}{4} c_I Y_{ij}{}^I \, T^{ab} R_{abk}{}^j(V) \,\varepsilon^{ki}
  \nonumber \\[.5ex] 
  &&{}
  + c_I\sigma^I\Big[ 
     \frac{1}{64} R_{ab}{}^{cd}(M)\,R_{cd}{}^{ab}(M) +\frac{1}{96}
  R_{ab j}{}^i(V) \,R^{ab}{}_i{}^j(V) \Big]  \nonumber\\[.5ex]
  &&{}
   -\frac{1}{128} {i}e^{-1}
  \,\varepsilon^{\mu\nu\rho\sigma\tau}\,c_IW_\mu{}^I\left[  
  R_{\nu\rho}{}^{ab}(M)\,R_{\sigma\tau ab}(M)+ \frac{1}{3}
  R_{\nu\rho j}{}^i(V) \,R_{\sigma\tau i}{}^j(V)\right] 
  \nonumber\\[.5ex] 
  &&{}  + \frac{3}{16} c_I\big(10\,\sigma^I \,T_{ab}- F_{ab}{}^I\big)\,
  R(M)_{cd}{}^{ab}\,T^{cd} \nonumber\\[.5ex]
  &&{} 
  + c_I\sigma^I\Big[ 3\,T^{ab} {D}^c {D}_aT_{bc} -\frac{3}{2}
  \big( {D}_aT_{bc}\big)^2 
  + \frac{3}{2} {D}_cT_{ab}\, {D}^aT^{cb}
  - \mathcal{R}_{ab}(T^{ac} T^b{}_c - \frac{1}{2} \eta^{ab} T^2) \Big]  
   \nonumber\\[.5ex]
   &&{} 
   + c_I\sigma^I\Big[
   \frac{8}{3} D^2 + 8\, T^2\,D - \frac{33}{8} (T^2)^2 + \frac{81}{2}
   (T^{ac}T_{bc})^2 \Big]    \nonumber\\[.5ex]
  &&{}
   - c_IF_{ab}{}^I\Big[T^{ab}\,D +\frac{3}{8}T^{ab} \,T^2 - \frac{9}{2} \, 
  T^{ac}T_{cd}T^{db} \Big] \nonumber\\[.5ex]
  &&{}
  +  \frac{3}{4} {i} \,\varepsilon^{abcde}\Big[c_I F_{ab}{}^I
  \big(T_{cf} {D}^fT_{de} +\frac{3}{2} T_{cf}
   {D}_dT_e{}^f\big)
   -  3\,c_I\sigma^I T_{ab} T_{cd}\, {D}^fT_{fe}\Big]\,.  
\end{eqnarray}

The constants $c_I$ encode the couplings of the higher derivative terms. The symbol $e$ denotes $e=\text{det}(e^a_{\mu})=\sqrt{-g}$.

Note that $\mathcal{R}$ and $\mathcal{R}_{ab}$ are respectively the Ricci scalar and tensor \footnote{In \cite{deWit:2009de} the authors use a different convention for the spin connection. This results in a sign flip for the curvature tensors.} while $R_{abcd}$ is the superconformal Weyl tensor. Other conventions can be found in the Appendix.

In the following we show how to obtain on-shell \textit{Poincar\'{e}} supergravity by integrating out the auxiliary fields.
The equation of motion for the auxiliary field $D$ is
\begin{equation}
 \frac{16}{3} c_I\sigma^I D + c_I(8\,\sigma^I T_{ab}- F_{ab}{}^I)
  \,T^{ab} +4\,C(\sigma) + 2\, \chi =0\,. 
\end{equation}which on the attractor background (\ref{attractorBackg}) reduces to
\begin{equation}\label{hyperScalar}
\chi= - 2\,C(\sigma) - 2\, c_I\sigma^I\,T^2 \,.
\end{equation}

For simplicity consider the theory with a unique vector multiplet without higher derivative corrections, that is, $c_I=0$. The function $C(\sigma)$ becomes $C(\sigma)=\sigma^3$. The gauge theory sector of the lagrangian, composed of a scalar $\sigma$, a vector $W_{\mu}$ and auxiliary fields $Y^{ij}$ becomes, after reintroducing the fermion fields, invariant under rigid superconformal transformations. Due to scale invariance we fix the scalar to a constant. If we further use the attractor equations $Y^{ij}=0$ and $T_{ab}=(4\sigma)^{-1}F_{ab}$ (\ref{attractorBackg}) we obtain
\begin{equation}
  \label{eq:pure-sg}
  8\pi^2\mathcal{L} = - \frac{1}{2} \sigma^3 \mathcal{R}
  -\frac{3}{8}\sigma \,F_{\mu\nu}F^{\mu\nu} -
  \frac{1}{8} i\,e^{-1} \varepsilon^{\mu\nu\rho\sigma\tau} W_\mu 
  F_{\nu\rho} F_{\sigma\tau} \,,
\end{equation}which upon including the gravitino field, is equal to the Lagrangian of pure five-dimensional supergravity. The Newton's constant is identified with $G_N=\sigma^{-3}$ so that the Ricci scalar appears with the canonical prefactor $(16\pi G_N)^{-1}$.

\subsection{BPS attractor equations and near horizon geometry}\label{on-shell solt}
In this section we present the attractor field configuration that preserves full supersymmetry. The analysis is completely off-shell and therefore it does not depend on the specific higher derivative corrections the theory may contain. To fully determine the black hole attractor background these equations must be supplemented with the values of the charges which depend on details of the higher derivative corrections. For further details we refer the reader to \cite{deWit:2009de}. 

Since ultimately we are interested in Poincar\'{e} supergravity we want to study the vanishing of the fermionic variations modded out by S-supersymmetry variations. This is achieved by constructing fermionic fields which are invariant under S-supersymmetry. This is basically the approach first outlined in \cite{LopesCardoso:2000qm}. The solutions are
\begin{equation}\label{attractorBackg}
\begin{array}{rcl}
&&\textbf{Vector mtpl. :}\\
&&\partial_{\mu}{\sigma^I} = 0,\\
&&F_{ab}^I = 4\sigma^I T_{ab}\\
&&Y^{ij} = 0
\end{array}
\qquad
\begin{array}{rcl}
&&\textbf{Weyl mtpl. :}\\
&&D_{[a}T_{bc]} = 0\\
&&D_{b}T^{ba} = i\varepsilon^{abcde}T_{bc}T_{de}\\
&&R_{\mu\nu i}\,^j(V) = 0\\
&&D = 0
\end{array}
\qquad
\begin{array}{lcr}
&&\textbf{Hyper mtpl. :}\\
&&\partial_{\mu}\chi = 0\\
&&D_{\mu}A_i^{\alpha} = 0\\
&& \chi \propto C(\sigma)
\end{array}
\end{equation}where $R_{\mu\nu i}\,^j(V)$ is the $SU(2)$ R-symmetry field strength $R_{\mu\nu i}\,^j(V)=2\partial_{[\mu}V_{\nu]i}\,^j-V_{[\mu i}\,^kV_{\nu] k}\,^j$.

The geometry has the form of a circle non-trivially fibered over $AdS_2\times S^2$
\begin{eqnarray}
\label{linelement}
ds^2=\frac{1}{16v^2}\left(-(r^2-1)dt^2+\frac{dr^2}{r^2-1}+d\theta^2+\sin(\theta)^2d\varphi^2\right)+e^{2g}(d\psi+B)^2,
\end{eqnarray}with fiber
\begin{equation}\label{fiberB}
B=\frac{1}{4v^2}e^{-g}\big(T_{23}(r-1) dt+T_{01}\cos(\theta)d\varphi \big)
\end{equation}
The size of $AdS_2\times S^2$ is determined via the condition 
\begin{equation}\label{vTT eq}
v^2=(T_{01})^2+(T_{23})^2
\end{equation}where $T_{01},\,T_{23}$ are the only non-vanishing components of $T_{ab}$, where $(0,1,2,3)$ are the local Lorentz indices.

For $T_{01}\neq 0$ the line element (\ref{linelement}) can be rewritten for $r\gg 1$ as
\begin{eqnarray}
  ds^2&=&- \frac{\rho^4}{16\,v^2}\left(\frac{T_{01}}{v}\,dt -
  \frac{T_{23}}{v\,\rho^2}\Big(\cos\theta\,d\varphi +
  \frac{1}{p^0}\,d\psi \Big)\right)^2 \nonumber\\
  &&{}\label{linelement 2}
  + \frac{1}{4\,v^2\rho^2}\left(d\rho^2 +
  \frac{\rho^2}{4}\Big(d\theta^2  + d\varphi^2
  +\frac{1}{(p^0)^2} \,d\psi^2  + \frac{2}{p^0} 
  \,\cos\theta\,d\varphi\,d\psi\Big) \right) \;, 
\end{eqnarray}
with 
\begin{equation}
\rho=\sqrt{r},\;p^0=\frac{e^{-g}}{4v^2}T_{01}
\end{equation}Up to the conformal factor $(4 v^2\rho^2)^{-1}$, the second term in the line element (\ref{linelement 2}) is diffeomorphic to flat space. However for $p^0\neq 1$ we have a conical singularity at the origin. Requiring smoothness of the solution we fix $p^0=1$ by imposing the condition
\begin{equation}
\frac{e^{-g}}{4v^2}T_{01}=1.
\end{equation}

Since the theory is scale invariant we set $v=1/4$ for convenience. The geometry is left with only one parameter $\beta\in [0,\pi/2[$ defined via the equation (\ref{vTT eq}) by setting
\begin{equation}\label{on-shell T}
T_{01}=\frac{1}{4}\cos(\beta),\,T_{23}=\frac{1}{4}\sin(\beta).
\end{equation}The line element (\ref{linelement}) becomes
\begin{equation}\label{linelementBeta}
ds^2=-(r^2-1)dt^2+\frac{dr^2}{r^2-1}+d\theta^2+\sin(\theta)^2d\varphi^2+\cos(\beta)^2\big(d\psi+\cos(\theta)d\varphi+\tan(\beta)(r-1)dt\big)^2
\end{equation}This is the near horizon geometry of a rotating black hole with angular momentum proportional to $J\propto \sin(\beta)$. The limiting case $\beta=\pi/2$ or $T_{01}=0$ has line element
\begin{equation}
ds^2=-(r^2-1)dt^2+\frac{dr^2}{r^2-1}+\left(d\psi+(r-1)dt\right)^2+ds^2(S^2).
\end{equation}The first three terms describe a local $AdS_3$. So effectively, we have the space $AdS_3\times S^2$. If we insist on the identification $\psi\sim \psi +4\pi$ we have the near horizon geometry of a black ring, while for noncompact $\psi$ we have an infinite black string. In this work we will be interested only in the case of a rotating black hole. The $AdS_3$ case, which is very interesting, will be postponed for a future work.

In summary, we have a one parameter family of geometries, which are locally $AdS_2\times S^2\times S^1$, that interpolate between the non-rotating black hole with near horizon geometry $AdS_2\times S^3$ and the black ring/string with near horizon geometry $AdS_3\times S^2$ \cite{LozanoTellechea:2002pn}.

The euclidean version of this configuration follows from a standard Wick rotation $t=-i\tau$ together with $T_{23}\rightarrow -i T^E_{23}$. This is equivalent to the transformation $\beta\rightarrow -i\alpha$. Definitions (\ref{on-shell T}) become
\begin{equation}
T_{01}=\frac{1}{4}\cosh(\alpha),\,T^E_{23}=\frac{1}{4}\sinh(\alpha).
\end{equation}and the line element (\ref{linelementBeta}) becomes
\begin{equation}
ds^2=(r^2-1)d\tau^2+\frac{dr^2}{r^2-1}+d\theta^2+\sin(\theta)^2d\varphi^2+\cosh(\alpha)^2\Big(d\psi+\cos(\theta)d\varphi-\tanh(\alpha)(r-1)d\tau\Big)^2
\end{equation}with $\tau \in [0,2\pi]$. In the rest of the paper we will use hyperbolic coordinates $r=\cosh(\eta)$ which appear to be more convenient. The line element is now
\begin{equation}\label{linelement Euclid}
ds^2=\sinh(\eta)^2d\tau^2+d\eta^2+d\theta^2+\sin(\theta)^2d\varphi^2+\cosh(\alpha)^2\Big(d\psi+\cos(\theta)d\varphi-\tanh(\alpha)(\cosh(\eta)-1)d\tau\Big)^2
\end{equation}with conformal boundary at $\eta\rightarrow \infty$, while the fiber becomes
\begin{equation}
B=\cos(\theta)d\varphi-\tanh(\alpha)(\cosh(\eta)-1)d\tau.
\end{equation}From a four dimensional point of view this corresponds to a $U(1)$ gauge field with electric and magnetic components. In the rest of the paper we use the vielbein basis
\begin{equation}
e^i=\Big(\sinh(\eta)d\tau,\,d\eta,\sin(\theta)d\varphi,\,d\theta,\,\cosh(\alpha)(d\psi+B)\Big),\,i=0,1,2,3,4
\end{equation}and inverse vielbein $E_i=E_i^{a}dx_{a}$
\begin{equation}
E_i^{\mu}=(e^{i}_{\mu})^{-1},\,E_i^{\psi}=-E^{\mu}_iB_{\mu},\,E^{\psi}_4=\cosh(\alpha)^{-1}
\end{equation}

 To find the gauge field attractor configuration we decompose the gauge field into a four  and five dimensional components  $A^{4d}$ and $\chi$ respectively,
\begin{equation}
A^{5d}=A^{4d}+\chi \cosh(\alpha)(d\psi+B),
\end{equation}respecting the symmetries of the near horizon geometry (\ref{linelement Euclid}). The field strength $F^{5d}$ has components
\begin{eqnarray}
 &&F^{(5d)}_{ij}=F^4_{ij}+\chi \cosh(\alpha)F(B)_{ij}\\
&& F^{(5d)}_{4i}=0
\end{eqnarray}for constant $\alpha,\chi$, where $F^4_{ij}=E^{\mu}_iE^{\nu}_jF_{\mu\nu}(A^4)$ with $\mu,\nu$ four dimensional indices. From the attractor equations 
\begin{equation}
F_{ab}=4\sigma T_{ab}\nonumber
\end{equation}we derive
\begin{equation}
F_{01}=-i\sigma \cosh(\alpha),\;F_{23}=-i\sigma \sinh(\alpha)
\end{equation}from which we can construct the five-dimensional gauge field
\begin{eqnarray}\label{attractor sols: gauge field}
A^{5d}=ie(\cosh(\eta)-1)d\tau+\chi \cosh(\alpha)(d\psi+B)
 \end{eqnarray}with
 \begin{equation}
 e=\frac{\sigma^*}{\cosh(\alpha)}\;\chi=-i\sigma^*\tanh(\alpha).
 \end{equation}Note that $d\psi+\cos(\theta)d\varphi$ is a globally defined form on $S^3$ {}\footnote{To see this consider complex coordinates $z_1,z_2$ on $\mathbb{C}^2$ with the parametrization $z_1=\cos(\theta/2)e^{i(\psi+\varphi)/2}$, $z_2=\sin(\theta/2)e^{i(\psi-\varphi)/2}$, then $d\psi+\cos(\theta)d\varphi=\text{Im}(z_1d\bar{z_1}+z_2d\bar{z_2})$.}. Note that the action contains Chern-Simons terms and therefore we want to use globally defined forms. For instance in the case of black rings the topology of the horizon is now $S^1\times S^2$ and so we can put magnetic flux on the $S^2$. This requires a careful treatment of the gauge fields as explained in \cite{deWit:2009de}. The field strength becomes
 \begin{equation}
 F^{5d}=-i\sigma^*\cosh(\alpha)\sinh(\eta)d\tau \wedge d\eta -i\sigma^*\sinh(\alpha)\sin(\theta)d\varphi\wedge d\theta
 \end{equation}

\subsection{Entropy, angular momentum and electric charges}
To compute the entropy, electric charges and angular momentum we can use the usual Noether procedure. Since the theory contains Chern-Simons terms this requires a careful treatment of the various fields. We present the results of \cite{deWit:2009de}. 
\begin{description}
\item[Entropy] the entropy follows from the 3-integral over the horizon of the Noether potential associated with space-time diffeomorphisms. This is particularly difficult due to the higher derivative terms and subtle due to the presence of Chern-Simons terms. Nevertheless we will see later in section \S\ref{Sec Local in 5D} that the entropy comes out naturally by computing the renormalized entropy function on the attractor background. Its value is
\begin{equation}
S_{BH}=\frac{\pi e^g}{4v^2}\Big(C(\sigma_*)+4c_I\sigma^I_*T_{23}^2\Big)
\end{equation}where $\sigma_*$ denotes the attractor value of the scalar.
\item[Angular momentum] If we consider the Noether potential associated with the Killing vector $\partial/\partial_{\psi}$ we compute the angular momentum
\begin{equation}\label{angMomentum}
J=\frac{T_{23}e^{2g}}{T_{01}^2}\Big(C(\sigma_*)-4c_I\sigma^I_*T_{01}^2\Big)
\end{equation}

\item[Electric charges] The electric charges are determined by considering the Noether potential associated with abelian gauge transformations. They are given by
\begin{equation}
q_I=\frac{3e^g}{2T_{01}}[C_{IJK}\sigma^J_*\sigma^K_*-c_IT_{01}^2]
\end{equation}
\end{description}

In the euclidean theory, entropy, angular momentum and electric charges become respectively 
\begin{eqnarray}\label{Euclidean charges}
S_{BH}&=&4\pi \cosh(\alpha)\Big(C(\sigma)-\frac{1}{4}c_I\sigma^I\sinh(\alpha)^2\Big),\\
J&=&4i\sinh(\alpha)\Big(C(\sigma)-\frac{1}{4}c_I\sigma^I\cosh(\alpha)^2\Big),\\
q_I&=&6\Big(C_{IJK}\sigma^J\sigma^K-\frac{1}{16}c_I\cosh(\alpha)^2\Big).
\end{eqnarray} Note that the angular momentum carries an imaginary factor $i$. This is a consequence of the fact that the fiber $B$ becomes real in the euclidean theory while the other gauge fields become imaginary. Once the charges and angular momentum are specified the attractor background is fully determined.

\section{Localization of $5D$ supergravity on $AdS_2\times S^2\ltimes S^1$}\label{Sec Local in 5D}

To proceed with equivariant localization we need two basic ingredients. Firstly, we need a fermionic symmetry $\delta$ that can be used to define a twisted de Rham operator in the sense of (\ref{delta squared}). Secondly, to show localization of supergravity on asymptotic $AdS_2\times S^2\ltimes S^1$ background we have to ensure that the integrand is equivariantly closed, that is, invariant under the fermionic symmetry $\delta$. Even though for supersymmetric theories defined on compact manifolds the last condition is satisfied by construction \footnote{We assume the measure to be invariant under the fermionic symmetry}, for spaces with boundaries, which is the case, the action functional is equivariantly closed only up to boundary terms. A different but equivalent way to understand this is to observe that the equations of motion for the gauge fields are not obeyed at the boundary as they carry non-normalizable components. To cure the theory we add appropriate boundary terms that compensate for the anomalous transformations. These terms take the form of five dimensional Wilson lines.

\subsection{Boundary terms and Wilson lines}\label{Bnd terms}
The lack of $\delta$ invariance can be restored by adding appropriate boundary terms. For the problem in hands it is enough to consider $\delta$ variations of the fields that carry non-normalizable components at the boundary, that is, the gauge fields and the fiber $B$ (\ref{fiberB}).

For ilustrative purposes consider the model with a single gauge field and Lagrangian
\begin{equation}
 \mathcal{L}= F\wedge \star F+\alpha F\wedge \star T+\beta A\wedge F\wedge F
\end{equation}in which the two derivative sector of our theory fits naturally. A variation of $\mathcal{L}$ under $A+\delta A$ \footnote{Note that $\delta A$ is an anticommuting field. However this analysis is independent of the commuting character of the variation.} gives a bulk plus boundary terms
\begin{equation}
\delta \mathcal{L}=2\delta A\wedge d\star F+\alpha\delta A\wedge d\star T+3\beta\delta A\wedge F\wedge F+2 d(\delta A\wedge \star F)+\alpha d(\delta A\wedge \star T)+2\beta d(\delta A\wedge A\wedge F)
\end{equation}at "order" $\delta A$. The last three terms being total derivatives will give contributions at the boundary. Consequently, to make the action $\delta$ invariant we add the boundary terms
\begin{equation}
S_{\text{bnd}}=-\int 2 d(\delta A\wedge \star F)+\alpha d(\delta A\wedge \star T)+2\beta d(\delta A\wedge A\wedge F).
\end{equation} To compute these boundary terms we use the attractor values of the fields. Paying careful attention to the orientation chosen\footnote{We have chosen $\int d\tau\wedge d\eta\wedge d\varphi\wedge d\theta\wedge d\psi=\int d\tau d\eta d\varphi d\theta d\psi$} we can show that the boundary term simplifies to
\begin{equation}\label{S_bnd}
 S_{\text{bnd}}=\tilde{Q}\oint_{S_{\tau}} A+g\oint_{S_{\psi}}\sqrt{h(r_0)}A
\end{equation}with $\tilde{Q}$ the flux
\begin{equation}
 \tilde{Q}=\int_{S^2\times S_{\psi}}\Big(2 \star F+\alpha \star T+2\beta A\wedge F\Big)\lvert_{on-shell},
\end{equation}and 
\begin{eqnarray}
&&g=\lim_{r_0\rightarrow \infty}\frac{1}{\sqrt{h(r_0)}}\int_{S_{\tau}\times S^2} \Big(2\star F+\alpha \star T+2\beta  A\wedge F\Big)\lvert_{on-shell}
\end{eqnarray}$h(r_0)$ is the induced metric at the boundary of $AdS_2$ with cutoff $r_0$ (\ref{RenEntropyFnct}). For the attractor solution  (\ref{attractor sols: gauge field} )we have
\begin{eqnarray}
\tilde{Q}_I&=&\frac{1}{2}i C_{IJK}\sigma_*^J\sigma_*^K(5+\cosh(2\alpha))\\
g_I&=&i\frac{1}{4}C_{IJK}\sigma^J_*\sigma^K_*\sinh(2\alpha)
\end{eqnarray}Even though the last term in (\ref{S_bnd}) cannot contribute to the on-shell renormalized action, since it's on-shell value is proportional to the cutoff $r_0$, it can contribute at the quantum level. In four dimensions $\tilde{Q}$ becomes the four dimensional charge and the boundary term (\ref{S_bnd}) reduces to a Wilson line insertion  on the thermal boundary as in \cite{Dabholkar:2010uh}.

The fiber also carries a non-normalizable component so we need to worry about possible new boundary terms. From a four dimensional point of view the fiber gives rise to an electric field. Small variations of the fiber generate total derivative terms that have to be compensated by boundary terms.  We compute these terms by reducing the theory to four dimensions  and studying the Maxwell kinetic term. For the second derivative lagrangian we compute 
\begin{equation}
S_{\text{fiber Bnd}}=-\tilde{J}\oint_{S_{\tau}} B,\,\text{with } \tilde{J}=\sinh(\alpha)\cosh(\alpha)^2C(\sigma_*)
\end{equation}

The discussion with higher derivative corrections follows exactly the same recipe, however the computation of the different Wilson lines is a hard task. For technical reasons, we decided to postpone for a future work the effects of higher derivative corrections in the computation of the quantum entropy.  Nonetheless we present here the analysis for the boundary terms. 

The electric Wilson lines get the additional contribution
\begin{equation}
\hat{Q}_I\oint_{S_{\tau}} A^I,\,\text{with }\hat{Q}_I=-i\frac{3}{32}\cosh(\alpha)^2\left(2+\sinh(\alpha)^2\right)c_I
\end{equation}while the Wilson line along the five dimensional circle acquires the additional term
\begin{equation}
\hat{g}_I\oint_{S_{\psi}}\sqrt{h(r_0)}A^I,\,\text{with }\hat{g}_I=-i\frac{3}{128}\sinh(2\alpha)\left(2+\sinh(\alpha)^2\right)c_I
\end{equation}
The computation of the boundary terms  for the fiber in the higher derivative lagrangian is not trivial. Special attention is needed to the term that couples the hypermultiplet scalar $\chi$ to the Ricci scalar
\begin{equation}
-\chi \frac{3}{16}\mathcal{R}
\end{equation}in the lagrangian (\ref{Lagr Hyper}). Its on-shell value (\ref{hyperScalar})
\begin{equation}
\chi|_{\text{on-shell}}=-2C(\sigma^*)-2c_I\sigma^{I*}T^2.
\end{equation}carries dependence on the constants $c_I$ and therefore needs to be taken into account.  After a tedious algebra and with the precious help of Mathematica we compute
\begin{equation}
-\hat{J}\oint_{S_{\tau}} B,\,\text{with }\hat{J}=-\frac{1}{128}c_I\sigma^{I*}\cosh(\alpha)^2(-49\sinh(\alpha)+3\sinh(3\alpha)).
\end{equation}With these boundary terms we can show that the on-shell renormalized action correctly reproduces the Wald's entropy. We will come back to this point later.

In summary, closeness of the path integrand under $\delta$ requires the supergravity action to be supplemented with additional boundary terms, that is,
\begin{equation}\label{newEntropyFnct}
iS_{sugra}+(\tilde{Q}_I+\hat{Q}_I)\oint_{S_{\tau}}A^I+(g_I+\hat{g}_I)\oint_{S_{\psi}}\sqrt{h(r_0)}A^I-(\tilde{J}+\hat{J})\oint_{S_{\tau}} B
\end{equation}where $S_{sugra}$ stands for supergravity action. Notice that neither $\tilde{Q}_I+\hat{Q}_I$ nor $\tilde{J}+\hat{J}$ match with the five dimensional electric charges and angular momentum respectively. However, as we shall see later on, the on-shell renormalized action correctly reproduces the entropy computed using the Noether methods.

\subsection{Localization}
So far we have not specified the fermionic symmetry $\delta$. As a matter of fact the analysis done in the previous section is equivalent to requiring that the equations of motion be obeyed also at the boundary \cite{Sen:2009vz}. This means that the boundary terms (\ref{newEntropyFnct}) are independent of the choice of $\delta$. 

For the problem in hands we have to consider a fermionic symmetry $\delta$ that is composed of a conventional Q and special S supersymmetries. In other words the fermionic symmetry is parametrized by spinors $\xi^i$ and $\eta^i$,
\begin{equation}\label{delta(xi,eta)}
\delta=\delta(\xi)+\delta(\eta).
\end{equation}Generally the $QV$ deformation  breaks most of the isometries of the problem because we choose to localize with a supercharge parametrized by a particular Killing spinor. That is, by choosing a Killing spinor $\xi$ we are in a sense gauge fixing part of the diffeomorphisms so this cannot be an exact deformation at least in supergravity. The susy transformations of section \S\ref{Sec: superconf} realize local superconformal symmetry and in general do not close to a circle action. Instead we will look at a region of the phase configuration space that realizes the equivariant algebra, that is, on which the fermionic symmetry closes to a compact symmetry modulo gauge transformations, in the sense that
\begin{equation}
\delta^2=\mathcal{L}_v+G
\end{equation}
 
 The superconformal transformations are very complicated and contain a large number of fields. To make things practical we consider the following ansatz for the five dimensional metric
\begin{equation}\label{metric in 5d}
ds^2=g_{\mu\nu}dx^{\mu}dx^{\nu}+\Phi^2(d\psi+B)^2
\end{equation}with $g_{\mu\nu}$ the four dimensional metric which asymptotes to $AdS_2\times S^2$. We assume that $g_{\mu\nu}$ is independent of the five dimensional coordinate  while both $\Phi$ and $B$ remain completely off-shell. 

Since the geometry is asymptotically a circle times $AdS_2 \times S^2$, the fermionic symmetry $\delta$ is expected to square at infinity to a Killing symmetry of $AdS_2\times S^2\times S^1$. From the $AdS_2$ point of view, there is a supercharge $Q$ in the near horizon superconformal algebra \cite{Banerjee:2009af} that squares to a compact symmetry, that is,
\begin{equation}\label{L-J}
Q^2=L_0-J
\end{equation}where $L_0$ generates rotations at the origin of $AdS_2$ and $J$, the two dimensional $R$-symmetry operator, generates azimuthal rotations on the sphere $S^2$. This was the supercharge used for localization in \cite{Dabholkar:2010uh,Gupta:2012cy}. However, while in \cite{Dabholkar:2010uh,Gupta:2012cy} the fermionic symmetry is generated by a Killing spinor of $AdS_2\times S^2$, here this is true only asymptotically. Because the localization equations allow for the five dimensional metric to fluctuate, as we will show,  both \emph{ $\xi^i$ and $\eta^i$ in (\ref{delta(xi,eta)}), the five dimensional susy parameters, will also have non-trivial profiles on the $AdS_2$ space while preserving the geometry at infinity}. We stress that this analysis is not exhaustive and a more general consideration could in principle be taken. For practical purposes we will only allow for a particular mode of the Killing spinor to fluctuate. We will show that the solutions are consistent with this assumption. 

In the case that the background along with the other fields in the Weyl multiplet are kept fixed to their on-shell values both $\xi$ and $\eta$ are determined by the vanishing of the susy variation of the gravitino. In the gauge $\eta=0$ there are eight independent Killing spinors. We refer the reader to the appendix (\ref{KillingSpinors}). The choice 
\begin{equation}\label{KillSpinor}
\xi=\left(\begin{array}{c}
\xi^1\\
\xi^2
\end{array}\right)=
\left(\begin{array}{c}
\xi_{++}\\
\xi_{--}
\end{array}\right)
\end{equation}generates the Killing vector 
\begin{equation}\label{Killing vec}
V^{\mu}\frac{\partial}{\partial x^{\mu}}=\frac{1}{2}\xi^{\dagger}\gamma^{\mu}\xi\frac{\partial}{\partial x^{\mu}}=-\frac{\partial}{\partial \tau}+\frac{\partial}{\partial \varphi}+\tanh(\alpha)\frac{\partial}{\partial \psi}.
\end{equation}and obeys a Majorana-symplectic condition. Moreover the Killing spinor on the space $AdS_2\times S^2\ltimes S^1$ is related to the Killing spinor on $AdS_2\times S^2$ used in \cite{Dabholkar:2010uh} by an $SO(1,1)$ transformation
\begin{equation}
\xi=e^{\frac{\alpha}{2} \gamma^4}\xi_{AdS_2\times S^2}
\end{equation}So indeed, in the case $\alpha=0$ we recover the symmetry (\ref{L-J}). Since we expect the parameter $\xi$ to not be completely fixed by the localization procedure we will use the ansatz
\begin{equation}\label{Killing spinor paramtrz}
\xi=e^{\frac{1}{2}\alpha(x,\psi)\gamma^4}\xi_{AdS_2\times S^2}
\end{equation}where $\alpha(x,\psi)$ is an arbitrary function which asymptotes to a constant $\alpha^*$.

The boundary conditions are as usual: we fix the non-normalizable modes and integrate the normalizable ones. This means
\begin{eqnarray}
&&\Phi=\cosh(\alpha)+\mathcal{O}(1/r)\nonumber\\
&&B_{\tau}=-\tanh(\alpha)(\cosh(\eta)-1)+\mathcal{O}(1),\,B_{\theta/\eta}=\mathcal{O}(1/r),\,B_{\varphi}=\cos(\theta)+\mathcal{O}(1/r)\nonumber\\
&&\sigma=\sigma_*+\mathcal{O}(1/r)\nonumber\\
&&A^{5d}_{\tau}=i\sigma^*\cosh(\alpha)(\cosh(\eta)-1)+\mathcal{O}(1),\,A_{\theta/\eta}=\mathcal{O}(1/r),\,A_{\varphi}=-i\sigma^*\sinh(\alpha)\cos(\theta)+\mathcal{O}(1/r)\nonumber\\
&&A^{5d}_{\psi}=\chi=-i\sigma^*\tanh(\alpha)+\mathcal{O}(1/r)\nonumber
\end{eqnarray}

\subsubsection{Localization in non-rigid background}\label{Loceqs: non-rigid back}

The strategy that we pursue here is to find a nice truncation where the susy superconformal transformations generate an equivariant algebra in the sense that
\begin{equation}\label{delta^2}
\delta^2=\mathcal{L}_v+G
\end{equation}with $\mathcal{L}_v$ the Lie derivative and $G$ a gauge transformation. Due to the large number of fields and the complexity of the equations involved,  we consider an ansatz for the metric and the susy parameter $\xi$. By computing second variations of the susy transformations we have to impose some constraints on the fields such that (\ref{delta^2}) is satisfied.


In the following we discuss the fermionic variations of both the Weyl and Vector multiplet fields. This discussion is tightly correlated with the off-shell reduction of five dimensional supergravity studied in \cite{Banerjee:2011ts}. However, since we are working with Euclidean space there are important details that have to be reconsidered. In this discussion we will try to keep the fields as much off-shell  as possible.

\paragraph{Weyl multiplet:}

We assume a Kaluza-Klein decomposition of the gravitino fields
\begin{equation}\label{KK reduction}
\psi^i_M=\left(\begin{array}{c}
\psi^i_{\mu}+B_{\mu}\tilde{\psi}^i\\
\psi^i=\tilde{\psi}^i+\check{\psi}^i
\end{array}\right)
\end{equation}where $M$ and $\mu$ are five and four dimensional coordinate indices, respectively, and $i$ is the $SU(2)$ R-symmetry index. The field $B_{\mu}$ is the fiber in  the metric (\ref{metric in 5d}). Since we are considering an off-shell fiber we need to keep the term $\check{\psi}^i$. For a standard reduction, that is, when the fiber does not depend on the five dimensional coordinate this term is zero. Similarly we write the decomposition of the other fields in the Weyl multiplet:
\begin{equation}
\mathcal{V}_{M\,j}^{\quad i}=\left(\begin{array}{c}
\mathcal{V}^{\quad i}_{\mu\,j}+B_{\mu}\mathcal{V}^{\;\; i}_j\\
\mathcal{V}^{\;\; i}_j
\end{array}\right),\,
T_{MN}=\left(\begin{array}{c}
T_{\mu\nu}\\
T_{4\mu}=A_{\mu}
\end{array}\right)
\end{equation}Note that the field $T$ goes to $-iT$ in Euclidean space. 

Since we are decomposing the five dimensional fields in four dimensional ones we need to ensure that certain gauge conditions of the superconformal algebra are still preserved. As explained in \cite{Banerjee:2011ts} this accounts to include additional Lorentz and special conformal compensating transformations of the fields. 

After some algebra, which is explained in the appendix (\ref{app susy variations Weyl}), the susy transformations of the fields $\Phi$, $B$ and $e^i_{\mu}$, the four dimensional vielbein, become
\begin{eqnarray}
&&\delta e^i_{\mu}=\frac{1}{2}\xi^{\dagger}\gamma^i\psi_{\mu}\nonumber\\
&&\delta \Phi=\frac{1}{2}\xi^{\dagger}\gamma^4\psi\nonumber\\
&&\delta B_{\mu}=\frac{1}{2}\Phi^{-2}\xi^{\dagger}\tilde{\gamma}_{\mu}\psi+\frac{1}{2}\Phi^{-1}\xi^{\dagger}\gamma^4\psi_{\mu}\nonumber
\end{eqnarray}which now look like the four dimensional susy transformations with $\tilde{\gamma}_{\mu}=e^i_{\mu}\gamma_i$.

We would like to stress that most of the susy transformations that we will be considering here are not supercovariant. Additional fermionic terms have to be added to the susy transformations of the fermionic fields. However since our main interest is on the solutions to the localization equations we can focus on the bosonic contributions only.

The action of $\delta^2$ on the fields $\Phi$ and $B$ becomes
\begin{eqnarray}
\delta^2\Phi &=&\frac{1}{2}V^{M}\partial_{M}\Phi+\frac{1}{2}\Phi\partial_{\psi}(\Phi^{-1}V^5)-\frac{1}{2}\partial_{\psi}(\Phi B_i)V^{i}\nonumber\\
\delta^2B_{\mu}&=&\frac{1}{2}V^{\nu}F_{\nu\mu}(B)+\Phi^{-1}V^{j}\partial_{\psi}(\Phi B_{[j})B_{\mu]}+\frac{1}{2}\Phi^{-2}V^5\partial_{\psi}(\Phi B_{\mu})-\frac{1}{4}\braket{\xi}{\xi}\partial_{\psi}\alpha B_{\mu}\nonumber\\
&&+\frac{1}{2}\Phi^{-1}(*\tilde{T})_{\mu j}V^{j}+\frac{1}{2}V^5\partial_{\mu}\Phi^{-1}-\frac{3}{2}\Phi^{-1}\braket{\xi}{\xi}A_{\mu}+\frac{1}{2}\Phi^{-1}\xi^{\dagger}\gamma^4\delta\psi_{\mu}\nonumber
\end{eqnarray}where we used a number of properties of the spinor $\xi$ as described in the appendix, and the vector $V^M$ is given by
\begin{equation}
V^M\partial_{M}=-\frac{\partial}{\partial\tau}+\frac{\partial}{\partial\varphi}+(-B_iV^i+\Phi^{-1}V^5)\frac{\partial}{\partial\psi}.
\end{equation}
 We see that both transformations do not obey the algebra (\ref{delta^2}). In $\delta^2\Phi$ only the first term corresponds to the action of $\mathcal{L}_v$ on $\Phi$ and therefore we need to impose
\begin{equation}
\Phi\partial_{\psi}(\Phi^{-1}V^5)-\partial_{\psi}(\Phi B_i)V^{i}=0.
\end{equation}For $\delta^2 B_{\mu}$ we have a bit more freedom because the fermionic symmetry can close up to gauge transformations since $B_{\mu}$ is now a four dimensional vector \footnote{Note that $\mathcal{L}_vA_{\mu}=(di_v+i_vd)A=i_vF(A)+d(i_vA)$.}. In addition $\delta^2 B_{\mu}$ contains non-linear terms acting on $B_{\mu}$ which have to vanish. This means that the terms $\Phi^{-1}V^{j}\partial_{\psi}(\Phi B_{[j})B_{\mu]}$, $\Phi^{-2}V^5\partial_{\psi}(\Phi B_i)$ and $\braket{\xi}{\xi}\partial_{\psi}\alpha B_{\mu}$ must vanish. To solve these constraints we need to impose 
\begin{equation}
\partial_{\psi}(B,\Phi,\alpha)=0,\nonumber
\end{equation}since we want to keep $B$, $\Phi$ and $\alpha$  as independent fields.

In addition there is in $\delta^2 B_{\mu}$  a  term proportional to the gravitino's variation $\delta\psi_{\mu}$. If we impose $\delta\psi_{\mu}=0$ together with $\psi_{\mu}=0$ we ensure that the vielbeins $e^i_{\mu}$ do not transform under $\delta^2$. In view of equivariant localization we want $\mathcal{L}_v$ to commute with the background such that in deforming the action by an exact term $\delta W[X,\Psi]$ we can pull out an action $\mathcal{L}_v$ through the functional and show that
\begin{equation}
\delta^2 W[X,\Psi]=\int \mathcal{L}_vW[X,\Psi]=\int \partial_v W=0\nonumber
\end{equation}This wouldn't be necessary if we knew how to construct a diffeomorphic and background independent exact deformation. In other words we need $\mathcal{L}_vg_{\mu\nu}=0$, that is, the vector $V^M$ should generate an isometry of the four dimensional metric. This can be achieved by imposing the four dimensional gravitino equation $\delta\psi_{\mu}=0$. 

The deformation $\delta W[X,\Psi]$ may depend on other parameters like the Killing spinor so we have to guaranty that they are also invariant under the flow generated by $v$. From our parametrization (\ref{Killing spinor paramtrz}) of the Killing spinor  this is not immediately true as $\alpha$ is an independent off-shell field.  In order to circumvent this problem we also need $\alpha$ to transform under supersymmetry as
\begin{equation}
\delta \alpha=\frac{1}{2}\Phi^{-1}\xi^{\dagger}\psi\nonumber
\end{equation}such that the action of $\delta^2$ on $\alpha$ is a translation along $v$, that is,
\begin{equation}
\delta^2\alpha=\frac{1}{2}V^M\partial_M\alpha=\mathcal{L}_v\alpha
\end{equation}The fact that $\alpha$ transforms under supersymmetry is natural from a four dimensional point of view where the theory has an additional $SO(1,1)$ R-symmetry. As a matter of fact, the field $\alpha$ joins the scalar $\Phi$ to form  a paracomplex scalar in the four dimensional theory \cite{Banerjee:2011ts,Cortes:2003zd}. As we show soon, the physical existence of $\alpha$ is offset by $A_i=T_{5i}$ via the condition $\delta\psi_{\mu}=0$ such that we are not adding additional degrees of freedom.

In view of these results the leftover expression\footnote{Observe that $\Phi^{-1}(*\tilde{T})$  contains a term linear in $F(B)$ so our reasoning is still valid in the sense that we need to keep this term in the susy variation.} in $\delta^2B_{\mu}$ must be a gauge transformation, that is,
\begin{equation}\label{eq gauge transf}
\Phi^{-1}(*\tilde{T})_{\mu j}V^{j}+V^5\partial_{\mu}\Phi^{-1}-3\Phi^{-1}\braket{\xi}{\xi}A_{\mu}=\partial_{\mu}\Lambda
\end{equation}  which is trivially satisfied after solving $\delta\psi_{\mu}=0$. 

From this analysis we conclude that in order for the fermionic transformation (\ref{delta(xi,eta)}) to square to a circle action on the Weyl multiplets fields we need the following conditions
\begin{equation}
\partial_{\psi}(B,\Phi,\alpha)=0,\,\,\delta \psi_{\mu}=0.
\end{equation}

 Let us now solve the four dimensional gravitino equation $\delta \psi_{\mu}=0$. It has been solved in \cite{Gupta:2012cy} to give $AdS_2\times S^2$ as the unique solution. However they considered the problem with Minkowski signature. The Euclidean case is not so different except that some fields have to be analytically continued to imaginary values. For instance we rewrite equation $\delta \psi_{\mu}=0$ as
\begin{equation}
(\frac{1}{2}\partial_{\mu}\alpha+3A_{\mu})\gamma^4\xi(0)^i+\frac{1}{4}(\bar{\omega}^{kl}_{\mu}-\bar{\omega}^{0kl}_{\mu})\gamma_{kl}\xi(0)^i+\frac{1}{4}(\hat{T}_{kl}-\hat{T}^{0}_{kl})\gamma^{kl}\gamma_{\mu}\xi(0)^i+\frac{1}{2}\mathcal{V}_{\mu j}^{\;\;\;i}\xi(0)^j=0
\end{equation}where we defined
\begin{equation}
\hat{T}_{ij}=\tilde{T}_{ij}\cosh(\alpha)-*\tilde{T}_{ij}\sinh(\alpha),
\end{equation}with $*T_{ij}=1/2\epsilon_{ijkl}T^{kl}$ and $\bar{\omega}^{0kl}_{\mu}$ is the spin connection of $AdS_2\times S^2$. At the on-shell level $\bar{\omega}^{kl}_{\mu}$ becomes $\bar{\omega}^{0kl}_{\mu}$. The field $\hat{T}^0$ corresponds to the on-shell value of $\hat{T}$ computed on $AdS_2\times S^2$ and has only one component $\hat{T}_{01}=1$. The spinor $\xi(0)$ denotes the Killing spinor of $AdS_2\times S^2$, that is, $\xi$ for $\alpha=0$.

 According to the authors of \cite{Gupta:2012cy} the only solution corresponds to $AdS_2\times S^2$, that is, for $\bar{\omega}^{kl}_{\mu}=\bar{\omega}^{0kl}_{\mu}$. Assuming $\mathcal{V}_{\mu j}^{\;\;\;i}=0$ we conclude that (see appendix for more details)
\begin{equation}
\hat{T}_{kl}=\hat{T}^{0}_{kl},\,\,A_{\mu}=-\frac{1}{6}\partial_{\mu}\alpha.
\end{equation}
With this result it is now easy to show that equation (\ref{eq gauge transf}) becomes a gauge transformation
\begin{equation}\label{gauge param 1}
\Phi^{-1}(*\tilde{T})_{\mu j}V^{j}+V^5\partial_{\mu}\Phi^{-1}-3\Phi^{-1}\braket{\xi}{\xi}A_{\mu}=\partial_{\mu}\left(J\cos(\theta)+H\cosh(\eta)\right)
\end{equation}with $J=\Phi^{-1}\cosh(\alpha)$ and $H=\Phi^{-1}\sinh(\alpha)$, as expected.
\newline

\paragraph{Vector multiplet:}

We now perform a similar analysis for the vector multiplet fields. The relevant susy transformations are
\begin{eqnarray}
&&\delta \sigma =\frac{1}{2}{i}{\xi}^{\dagger}\Omega\nonumber\\
&&\delta\Omega^i=-\frac{1}{4}F_{ab}\gamma^{ab}\xi^i-i\sigma T_{ab}\gamma^{ab}-\frac{i}{2}\gamma^a\partial_a\sigma \xi^i+Y^i_{\,\,j}\xi^j+\frac{1}{2}\sigma\eta^i\nonumber\\
&&\delta W_M= \frac{1}{2}\xi^{\dagger}\gamma_M\Omega- \frac{1}{2}{i}
  \sigma \xi^{\dagger} \psi_M \,. \label{delta W5dim}
\end{eqnarray}Along the standard Kaluza-Klein reduction we decompose the five-dimensional gauge field into a four dimensional component $W_{\mu}$ and a "scalar" $W_{\psi}$ as 
\begin{eqnarray}
&&W_{\psi}=\tilde{W}_{\psi}+\chi \Phi\\
&&W_{\mu}=\tilde{W}_{\mu}+\chi \Phi B_{\mu}
\end{eqnarray}Note that this vector is still completely off-shell. We have separated the five dimensional component in two parcels to put in evidence the Wilson line along $\psi$, that is, 
\begin{equation}
\mathcal{U}=\frac{1}{4\pi}\int d\psi W_{\psi}=\chi \Phi
\end{equation}
This ensures that $\partial_{\psi}(\chi \Phi)=0$ and $\int d\psi \tilde{W}_{\psi}=0$. However we do not make any other assumption about coordinate dependence of $\tilde{W}$. 

After some algebra it is possible to show that the action of  $\delta^2$ on the bosonic fields is
\begin{eqnarray}
&&\delta^2\sigma= \frac{1}{2}V^M\partial_M\sigma +\frac{1}{4}i\sigma \xi^{\dagger}\eta \nonumber\\
&&\delta^2 \tilde{W}_{\psi}=\frac{1}{2}V^{M}F_{M\psi}(\tilde{W})-\frac{i}{4}\partial_{\psi}(\sigma \braket{\xi}{\xi})\nonumber\\
&&\delta^2 \tilde{W}_{\mu}=\frac{1}{2}V^MF_{M\mu}(\tilde{W})-\frac{i}{4}\partial_{\mu}\Lambda \label{gauge param 2}\\
&&\delta^2\mathcal{U}=\frac{1}{2}V^M\partial_M\mathcal{U}\nonumber
\end{eqnarray}with the gauge parameter
\begin{eqnarray}
\Lambda=\big(\sigma\sinh(\alpha)+\chi\cosh(\alpha)\big)\cos(\theta)+\big(\sigma\cosh(\alpha)+\chi\sinh(\alpha)\big)\cosh(\eta)
\end{eqnarray}
From the first equation we conclude that $\eta$ must be "orthogonal" to $\xi$ in the sense that $\xi^{\dagger}\eta=0$. The choice for $\eta$ (\ref{eta})  trivially satisfies this condition
\begin{equation}
\xi^{\dagger}(2i\gamma^4 \Slash A \xi-\frac{i}{2}T_{kl}\gamma^{kl}\xi+\frac{i}{8}\mathcal{F}_{kl}\gamma^{kl}\gamma^4\xi)=0.
\end{equation}after using the property that $\xi^{\dagger}\gamma^{ab}\xi=0$.
The rest of the algebra is already in the form $\delta^2=(\mathcal{L}_v+G)$. 

\bigskip
Before proceeding with localization we make a brief summary of what we have done so far. Starting from an ansatz for the metric and Killing spinor and assuming a particular Kaluza-Klein reduction we computed the action of $\delta^2$ on the bosonic fields of both the Weyl and vector multiplets. Since we want $\delta^2$ to generate a circle flow this imposes additional constraints on the fields. From this analysis it results that
\begin{eqnarray}
&&\partial_{\psi}(B,\Phi,\alpha)=0,\,\,\delta \psi_{\mu}=0,\,\,\xi^{\dagger}\eta=0
\end{eqnarray}Note that we need to impose the  condition $\delta \psi_{\mu}=0$ before using the localization argument.
A similar analysis should be carried also for the fermionic fields, even though it should follow just by supersymmetry. We will skip this analysis and proceed to solving the localization equations which perturbative analysis only requires the supersymmetric transformations $\delta\Psi$.

On the field configuration space where $\delta^2$ acts equivariantly we can add an exact deformation of the form
\begin{equation}\label{exact deformation}
S\rightarrow S-t\int \delta\left((\delta\Psi)^{\dagger}\Psi\right)
\end{equation}where $\Psi$ runs through all the fermions of the theory and we keep fixed the four dimensional metric to $AdS_2\times S^2$ and $\psi_{\mu}=0$. From the analysis done before it is easy to show why that is an exact deformation. For any scalar functional $W[X,\Psi]$ the action of $\delta^2$ is simply
\begin{equation}
\delta^2 \int W[X,\Psi]=\int \mathcal{L}_vW[X,\Psi]=\int V^M\partial_MW[X,\Psi]
\end{equation}which vanishes after an integration by parts, that we can do because $\partial_MV^M=0$.
 The bosonic action that results from this deformation is $(\delta\Psi)^{\dagger}\delta\Psi$. So in the limit $t\rightarrow \infty$ we derive the localization equations
\begin{equation}
\delta \Psi=0.
\end{equation}

\subsubsection{Localization solutions}
In this section we solve the localization equations for the Weyl and Vector multiplet fields under the conditions derived in the previous section. 

\paragraph{Weyl multiplet:}
 Using the condition that $\partial_{\psi}(B,\Phi,\alpha)=0$,  equation $\delta\psi^i=0$ (\ref{locEq: psi}) becomes 
\begin{eqnarray}
&&F_{kl}(B)\gamma^{kl}\xi^i(0)+2\gamma^4\Slash{\partial}\big[\Phi^{-1}\cosh(\alpha)\big]\xi^i(0)
-2\Slash{\partial}\big[\Phi^{-1}\sinh(\alpha)\big]\xi^i(0)\nonumber\\
&&-\Phi^{-1}\big[\sinh(\alpha)\hat{T}_{kl}+\cosh(\alpha)(*\hat{T})_{kl}\big]\gamma^{kl}\xi^i(0)-2\Phi^{-2}\mathcal{V}^{\;\;i}_{\,j}\xi^j(0)=0
\end{eqnarray}
Notice that this equation has the form of a susy transformation of a vector multiplet fermion  except for a couple of imaginary factors. As a matter of fact this becomes the supersymmetry transformation of the four dimensional compensating vector multiplet fermion \cite{Banerjee:2011ts}. If we denote by $\hat{F}$ the field strength of the fluctuations $\delta B$ above the attractor value $B^*$, we have 
\begin{eqnarray}\label{locEq: psi-Wick-rotation}
&&\frac{1}{2}\hat{F}_{kl}\gamma^{kl}\xi^i(0)-H\gamma^{01}\xi^i(0)
-J\gamma^{23}\xi^i(0)+\gamma^4\Slash{\partial}J\xi^i(0)
-\Slash{\partial}H\xi^i(0)-\Phi^{-2}\mathcal{V}^{\;\;i}_{\,j}\xi^j(0)=0\nonumber\\
{}
\end{eqnarray}where we defined $H=\Phi^{-1}\sinh(\alpha)-\tanh(\alpha^*)$ and $J=\Phi^{-1}\cosh(\alpha)-1$. Both $H$ and $J$ vanish at the boundary. If the fields $\delta B$, $H$ and $J$ take real values this leads to an infinite number of solutions. To avoid this situation we perform a Wick rotation of the field $\delta B$ to $i\delta B^E$ \footnote{Analogously we could have considered the complexified version of $(\delta\psi)^{\dagger}$ in (\ref{exact deformation}) in the sense that we take $F(B)_{ij}$ to be a complex field with the reality condition that $F(B)_{ij}^{\dagger}=F(B^*)_{ij}-F(\delta B)_{ij}$, with $B^*$ the on-shell value. The resulting action would not be positive definite in this case. However this can be avoided by integrating over imaginary values of $\delta B_{\mu}$. }, which does not change the boundary conditions, and take the imaginary branch of $\mathcal{V}^{\;\;1}_{1}$, that is, $\mathcal{V}^{\;\;1}_{1}=-\mathcal{V}^{\;\;2}_{2}=iK$, with the other components zero. Remind that $\mathcal{V}^{\;\;i}_j$ is an antihermitian traceless matrix in the $SU(2)$ indices. Another possibility would be to Wick rotate both $H$ and $J$. However this would spoil the reality condition of $\alpha$ inducing important changes in the localization equations.  

Parametrizing the supersymmetry transformations in the form
\begin{equation}\label{parametrization susy}
\left(\begin{array}{c}
\delta \psi^1\\
\delta\psi^2
\end{array} \right)=\left(\begin{array}{c} 
- \frac{1}{2} \left(\tilde{F}_{ab}+i\tilde{G}_{ab}\right) \gamma^{ab} \xi^1
  -{i} \Slash{B} \xi^1 +\tilde{K}\xi^1+(\tilde{L}+\tilde{J})\xi^2\\
  - \frac{1}{2} \left(\tilde{F}_{ab}+i\tilde{G}_{ab}\right) \gamma^{ab} \xi^2
  -{i} \Slash{B} \xi^2 -\tilde{K}\xi^2+(\tilde{L}-\tilde{J})\xi^1
\end{array} \right)
\end{equation}we construct the bosonic part of the localization lagrangian as
\begin{eqnarray}\label{LocLagrangian Bosonic}
 \sum_{i=0,1}(\delta \psi^i)^{\dagger}\delta \psi^i &=&\frac{\braket{\xi}{\xi}}{4}\left(\tilde{F}_{mn}+\frac{1}{2}\epsilon_{mnpqr}\tilde{F}^{pq}\frac{V^r}{\braket{\xi}{\xi}}-\frac{J}{\braket{\xi}{\xi}}\Theta^2_{mn}\right)^2+\frac{1}{2\braket{\xi}{\xi}}(V^m\tilde{F}_{mn})^2+\nonumber\\
&&+\frac{\braket{\xi}{\xi}}{4}\left(\tilde{G}_{mn}+\frac{1}{2}\epsilon_{mnpqr}\tilde{G}^{pq}\frac{{V}^r}{\braket{\xi}{\xi}}+\frac{2}{\braket{\xi}{\xi}}B_{[m}V_{n]}-\frac{\tilde{K}}{\braket{\xi}{\xi}}\Theta^1_{mn}-\frac{L}{\braket{\xi}{\xi}}\Theta^3_{mn}\right)^2\nonumber\\
&&+\frac{\braket{\xi}{\xi}}{2}\left(\frac{V^m}{\braket{\xi}{\xi}}\tilde{G}_{mn}-B_n\right)^2+\frac{1}{2\braket{\xi}{\xi}}(B.V)^2
\end{eqnarray}More details about this construction can be found in the appendix \S\ref{App:locEqs}. We used the notations
\begin{equation}
\braket{\xi}{\xi}=\xi^{\dagger}\xi,\,V^a=\xi^{\dagger}\gamma^a\xi
\end{equation}such that the vector $V^a/\braket{\xi}{\xi}$ has unit norm. Since the bosonic lagrangian is written as a sum of squares the localization equations follow directly from the zero locus of each of these squares, that is,
\begin{eqnarray}
&&\tilde{F}_{mn}+\frac{1}{2}\epsilon_{mnpqr}\tilde{F}^{pq}\frac{V^r}{\braket{\xi}{\xi}}=0\label{eqLoc:1}\\
&&V^m\tilde{F}_{mn}=0 \label{eqLoc:2}\\
&&\tilde{G}_{mn}+\frac{1}{2}\epsilon_{mnpqr}\tilde{G}^{pq}\frac{{V}^r}{\braket{\xi}{\xi}}+\frac{2}{\braket{\xi}{\xi}}B_{[m}V_{n]}-\frac{\tilde{K}}{\braket{\xi}{\xi}}\Theta^1_{mn}=0\label{eqLoc:3}\\
&&\frac{V^m}{\braket{\xi}{\xi}}\tilde{G}_{mn}-B_n=0\label{eqLoc:4}\\
&&B.V=0\label{eqLoc:5}
\end{eqnarray}For the problem we are considering we need to set both $L$ and $J$ to zero. If it wasn't the case we could generate an infinite number of solutions to the localization equations. One possibility would be to consider a space-time dependent analytic continuation of the auxiliary fields as in \cite{Gupta:2012cy}.

Observe that some of the equations are not independent. For instance equation (\ref{eqLoc:2}) comes from equation (\ref{eqLoc:1}) after contraction with the vector $V$. Analogously, equation (\ref{eqLoc:4}) comes from contraction of (\ref{eqLoc:3}) with the vector $V$ after using equation (\ref{eqLoc:5}). 

Under the parametrization (\ref{parametrization susy}), we read
\begin{eqnarray}
&&\tilde{F}=-\hat{F}^E,\nonumber\\
&&B_i=\partial_i H,\, B_5=0\nonumber\\
&&\tilde{G}_{01}=H,\,\tilde{G}_{23}=J,\,\tilde{G}_{i4}=\partial_i J\nonumber\\
&&\tilde{K}=\Phi^{-2}K\nonumber
\end{eqnarray}
Equation (\ref{eqLoc:1}) is easily solved to give
\begin{equation}
\hat{F}=0\rightarrow B=B^*,\nonumber
\end{equation}that is, the fiber must be fixed to its on-shell value.
From (\ref{eqLoc:4}) we deduce 
\begin{equation}
V^i\partial_i J=0,\, \partial_i H-\frac{V^m}{\braket{\xi(0)}{\xi(0)}}\tilde{G}_{mi}=0.\nonumber
\end{equation}The second equation translates into the fact that the gauge parameter $\Lambda$ in (\ref{gauge param 1}) becomes a constant $C$ on the localization locus. The equations that follow from the "master equation" (\ref{eqLoc:3}) are
\begin{eqnarray}
&&\partial_{0,2}H=\partial_{0,2}J=0\nonumber\\
 && \cosh(\eta)H+\cos(\psi)J+\partial_1 H\sinh(\eta)+\partial_3 J\sin(\psi)-\tilde{K}=0 \label{eq1app}\nonumber\\
&& \cosh(\eta)J+\cos(\psi)H-\partial_1 J\sinh(\eta)-\partial_3 H\sin(\psi)-\tilde{K}\cos(\psi)\cosh(\eta)=0\label{eq2app}\nonumber\\
&& \partial_1 H \sin(\psi)=\partial_3 J \sinh(\eta)-\tilde{K}\sin(\psi)\sinh(\eta)\label{eq3app}\nonumber\\
&& \partial_3 H \sinh(\eta)=\partial_1 J \sin(\psi)\label{eq4app}\nonumber\\
{}\label{loc equations Dabholkar}
\end{eqnarray}which have been solved before in \cite{Dabholkar:2010uh,Gupta:2012cy}. The solutions are
\begin{equation}
H=\frac{C}{\cosh(\eta)},\,J=0,\,\tilde{K}=\frac{C}{\cosh(\eta)^2},
\end{equation}with $C$ an arbitrary constant (to be identified with the gauge parameter $\Lambda$, as pointed out just before). In terms of the fields $\Phi$ and $B$ this gives
\begin{equation}\label{loc solts Phi}
\Phi=\cosh(\alpha),\,\tanh(\alpha)=\tanh(\alpha^*)+\frac{C}{\cosh(\eta)},\,K=\cosh(\alpha)^2\frac{C}{\cosh(\eta)^2},\,B=B^*
\end{equation}From here we see that $C$ must be defined in the interval $[-1-\tanh(\alpha^*),1-\tanh(\alpha^*)]$.

We proceed with localization and consider the remaining fermionic fields in the Weyl multiplet. The field $\chi^i$ has an intricate susy transformation. Instead we use the results of \cite{Banerjee:2011ts}. The authors present its decomposition in terms of four dimensional fields
\begin{eqnarray}
\chi^i \big\vert_{4D}   =&
   8 \chi^i +\frac{1}{48} \gamma^{ab}{F}_{ab}\psi^i
   -\frac{3}{4} \,\Phi^{-1}\, T_{ab} \gamma^4 \gamma^{ab}\psi^i \,,
   \nonumber\\
   &+\frac{1}{4} \Phi\,\gamma_4 \Slash{D}( \Phi^{-2}\psi^i)
   - \frac{1}{2} \Phi^{-2} \mathcal{V}^i{}_j \psi^j
   + \frac{9}{4}  \Phi^{-1}\, A_a\gamma^a \psi^i \,,\nonumber
\end{eqnarray}Since both $\delta \chi^i$ and $\delta \psi^i$ vanish at the localization locus, this implies that $\delta \chi^i\big\vert_{4D}=0$. This has a much simpler expression we rather use 
\begin{eqnarray}\label{locEq:delta chi}
\delta \chi^i\big\vert_{4D}=&&\frac{1}{3}\gamma^{ij}\Slash{\nabla}\tilde{T}_{ij}\xi^i+2\gamma^{ij}\tilde{T}_{ij}\gamma^4\Slash{A}\xi^i+\tilde{D}\xi^i=0
\end{eqnarray}with $\tilde{D}$ defined as
\begin{eqnarray}
\tilde{D}=&&\,4\, D+ \frac{1}{4} \Phi\big(
  \nabla^a{\nabla}_a +\frac{1}{6}\mathcal{R}\big)\Phi^{-1} +
  \frac{3}{32}\Phi^2 {F}^{ab}{F}_{ab}\nonumber\\
  &&\,- \frac{3}{2} T^{ab}T_{ab} -3\, A^{a}A_a +\frac{1}{4}
  \Phi^{-2} \, \mathcal{V}_i{}^j\, \mathcal{V}_j{}^i \,.
\end{eqnarray}with $\mathcal{R}$ and $\nabla_a$ the four dimensional Ricci scalar and covariant derivative respectively. We have used the fact that $\tilde{\eta}=0$ and set $\mathcal{V}_{\mu}=0$. Note that $\tilde{D}$ vanishes on-shell. Substituting back the value of $A_{\mu}$ we find 
\begin{eqnarray}
\delta \chi^i\big\vert_{4D}=0&\Leftrightarrow & \frac{1}{3}\gamma^{kl}\Slash{\nabla}\hat{T}_{kl}\xi^i+\tilde{D}\xi^i=0\nonumber
\end{eqnarray}

Since the tensor $\hat{T}$ is covariantly constant, this implies
\begin{equation}
\tilde{D}=0.\nonumber
\end{equation} 

This finishes the analysis for the fields in the Weyl multiplet. 

\paragraph{Vector multiplet:}
 
As explained before we perfom an off-shell Kaluza Klein decompostion of the five dimensional gauge field as
\begin{equation}
A^{5d}=\tilde{W}+\mathcal{U}(d\psi+B)\nonumber
\end{equation}where $\mathcal{U}=\chi \Phi$ is the Wilson line along $\psi$. To obtain non-trivial solutions to the localization equations we have to analytically continue the field $\chi$ to imaginary values. This is consistent with the on-shell solution discussed in the section \S\ref{on-shell solt}. From a four dimensional point of view this is a consequence of the fact that $\mathcal{N}=2$ euclidean supersymmetry has $SO(1,1)$ R-symmetry \cite{Cortes:2003zd}, so that the vector multiplet scalars are real.

 The $\delta$ variation of the fermion $\Omega^i$ in the vector multiplet becomes
 \begin{equation}
\delta\Omega^i=-\frac{1}{4}F_{ab}\gamma^{ab}\xi^i-\frac{i}{4}\chi\Phi F(B)_{kl}\gamma^{kl}\xi^i-\frac{i}{2}\Phi^{-1}\gamma^k\gamma^4\partial_k(\chi\Phi)-i\sigma T_{ab}\gamma^{ab}-\frac{i}{2}\gamma^a\partial_a\sigma \xi^i+Y^i_{\,\,j}\xi^j+\frac{1}{2}\sigma\eta^i=0
\end{equation}where $a,b$ and $k,l,m$ are respectively five and four dimensional tangent space indices. With the help of equation $\delta\psi=0$ (\ref{locEq: psi-Wick-rotation}) and the fact that $\tilde{\eta}=0$ we rewrite the equation above as
\begin{eqnarray}
&&-\frac{1}{4}e^{-\frac{1}{2}\alpha\gamma_4}\hat{F}_{ab}\gamma^{ab}\xi^i-\frac{i}{2}\left[(\sigma +\gamma^4\chi)e^{\alpha\gamma^4}-(*)\right]\gamma^{01}\xi^i(0)-\frac{i}{2}\gamma^i\partial_{i}[(\sigma +\gamma^4\chi)e^{\alpha\gamma^4}-(*)]\xi^i(0)\nonumber\\
&&-\frac{i}{2}\gamma^4\partial_{4}\sigma \xi^i(0)+(Y^i_{\,\,j}+\frac{1}{2}\chi \Phi^{-1}\mathcal{V}^{Ei}_{\,\,j})\xi^j(0)=0
\end{eqnarray}where $\hat{F}$, which is taken to be real, denotes fluctuations of the gauge fields above the attractor background and $(*)$ is the on-shell value of $ (\sigma +\gamma^4\chi)e^{\alpha\gamma^4}$.  

The bosonic part of the localization lagrangian can be written again in the form (\ref{LocLagrangian Bosonic}). We read
\begin{eqnarray}
&&B_i=\frac{1}{2}\partial_{i}\big(\sigma\cosh(\alpha)+\chi\sinh(\alpha)\big),\,B_4=\frac{1}{2}\partial_{4}\sigma\\
&&\tilde{G}_{i4}=\frac{1}{2}\partial_{i}\big(\sigma\sinh(\alpha)+\chi\cosh(\alpha)\big),\\
&&\tilde{G}_{01}=\frac{1}{2}\left(\sigma \cosh(\alpha)+\chi\sinh(\alpha)-(*)\right),\,\tilde{G}_{23}=\frac{1}{2}\left(\sigma \sinh(\alpha)+\chi\cosh(\alpha)-(*)\right),\\
&&\tilde{K}=Y^1_{\,\,1}+\frac{1}{2}\chi \Phi^{-1}\mathcal{V}^{E1}_{\,\,1}
\end{eqnarray}
 It immediately follows from the localization equations
\begin{equation}
\frac{1}{\braket{\xi(0)}{\xi(0)}}V^i\tilde{G}_{i4}=B_4,\,B.V=0
\end{equation}and $V^i\partial_i\alpha=0$, that
\begin{equation}
\partial_{\psi}\sigma=\frac{\Phi}{2}V^i\partial_i\chi/\braket{\xi}{\xi}.\nonumber
\end{equation}The RHS of the equation does not depend on $\psi$. So in order to preserve the periodicity of $\sigma$ we must have
\begin{equation}
\partial_{\psi}\sigma=0,\,V^i\partial_i\chi=0.
\end{equation}We therefore conclude that $\sigma$ must live on $AdS_2\times S^2$. The remaining equations are analogous to the system (\ref{loc equations Dabholkar}) and can be solved to give
\begin{eqnarray}
&&\sigma\cosh(\alpha)+\chi\sinh(\alpha)=(*)+\frac{\tilde{C}}{\cosh(\eta)},\\
&&\sigma\sinh(\alpha)+\chi\cosh(\alpha)=(*)\\
&&Y^{1}_{\,\,1}+\frac{1}{2}\chi\Phi^{-1}\mathcal{V}^{E1}_{\,\,1}=\frac{\tilde{C}}{2\cosh(\eta)^2}
\end{eqnarray}with $\tilde{C}$ an arbitrary constant. We also observe that the equation 
\begin{equation}
V^m\tilde{T}_{m\mu}-\braket{\xi(0)}{\xi(0)}B_{\mu}=0
\end{equation}translates into the fact the the gauge transformation parameter $\Lambda$ in (\ref{gauge param 2}) is exactly the constant $\tilde{C}$. This tells us that we are integrating over constant gauge transformations. As a matter of fact this a common feature of localization. For instance, in BRST localization of SYM on $S^4$ \cite{Pestun:2007rz}, there is an auxiliary parameter $a_0$, coming from the Fadeev-Popov procedure, that gets identified with the adjoint scalar that is left unfixed by the localization equations. The field $a_0$ parametrizes gauge transformations since the BRST equivariant operator $Q$ squares to rotations plus gauge transformations parametrized by $a_0$. Its only on the localization locus that the constant value of the scalar gets identified with gauge transformations.

We now focus on the gauge sector. 

The localization equations imply the five dimensional "antiself-dual" Yang-Mills equation\footnote{This equation is known in the literature as the antiselfdual contact instanton equation and has appeared in many different contexts of five dimensional localization in gauge theories \cite{Kim:2012qf,Hosomichi:2012ek,Kallen:2012va}.}
\begin{equation}
\hat{F}_{mn}+\frac{1}{2}\epsilon_{mnpqr}\hat{F}^{pq}v^r=0
\end{equation}with $v^r$ the five dimensional unit vector ${V^r}/{\braket{\xi}{\xi}}$ and $\hat{F}=dA$, with $A$ denoting the fluctuations above the attractor value. It follows from contraction of this equation with $v^r$ that
\begin{equation}\label{ivF}
v^m\hat{F}_{mn}=0
\end{equation}Before trying to solve these equations note that the vector $V^{M}$ has components
\begin{equation}
\frac{1}{2}V^M\partial_M=-\frac{\partial}{\partial\tau}+\frac{\partial}{\partial\varphi}+\big[\tanh(\alpha^*)+C\big]\frac{\partial}{\partial\psi}
\end{equation}with $C$ the constant in (\ref{loc solts Phi}). It is therefore constant in the sense that $\partial_N V^M=0$. In a gauge where
\begin{equation}
V^MA_M=0,\nonumber
\end{equation}which we can choose because the Wilson lines have already been removed, equation (\ref{ivF}) translates into
\begin{equation}
v^M\partial_MA_N=0\nonumber.
\end{equation}In other words, in coordinates where $v^M\partial_M=\partial_z$, this equation is simply the statement that $A$ does not depend on the coordinate $z$, while the gauge condition becomes equivalent to $A_z=0$. 

It follows that the five dimensional "anti-selfdual" YM equation reduces to
\begin{equation}
F_{mn}+\frac{1}{2}\epsilon_{mnpqz}F^{pq}=0
\end{equation}which is just the anti-selfdual YM equation in the four dimensional space transverse to $v$. This equation is also known as contact instanton equation \cite{Kallen:2012va}. The kernel of  $k=v_rdx^r$ defines a four dimensional "orthogonal" space $\mathcal{M}$ via $k(\xi)=0$ with $\xi\in T\mathcal{M}$  \cite{New1}. In the absence of contact instantons the five dimensional "anti-selfdual" YM equation implies
\begin{equation}
d*F=0|_{\mathcal{M}}\Leftrightarrow A=0\text{ mod gauge transf.}
\end{equation} unless $\mathcal{M}$ contains non-trivial one-cycles. Without entering in details about the topological properties of $\mathcal{M}$ we will assume that this is the case.

\subsection{Quantum Entropy function}

Our task now is to compute the action on the localization solutions. As discussed in section \S\ref{QEF} the action suffers from IR divergences due to the infinite volume of $AdS_2$. However they can be renormalized systematically by introducing appropriate local boundary counter terms. 

In \cite{Banerjee:2011ts} the authors performed not only the Kaluza Klein reduction of five dimensional off-shell multiplets but they have rewritten part of the five dimensional action in terms of four dimensional fields. Their results are very interesting. They observe that the two derivative lagrangian together with part of the higher derivative corrections can be rewritten in terms of four dimensional chiral superspace invariant terms, usually called F-terms. This part of the action can be written in terms of the holomorphic prepotential function
\begin{equation}
F(X,\hat{A})=a C_{IJK}\frac{X^IX^JX^K}{X^0}+b \hat{A}\frac{c_I X^I}{X^0}
\end{equation}with $a,b$ some numerical constants. This type of lagrangian falls in the class of theories reviewed in \cite{Mohaupt:2000mj} relevant for BPS black holes in $\mathcal{N}=2$ supergravity and more recently in the case of localization of supergravity in $AdS_2\times S^2$ \cite{Dabholkar:2010uh}. Interestingly though some of the higher derivative terms give unexpected contributions in four dimensions.  For instance they give rise to Gauss-Bonnet type of corrections in four dimensions which have never been written in $\mathcal{N}=2$ supergravity. Other terms can be written as integrals over the full superspace usually known as D-terms. This class of terms was extensively analyzed in \cite{deWit:2010za}. Their analysis however is not fully complete as there are a number of terms whose reduction can be ambiguous because of integration by parts. On the other hand our analysis in section \S\ref{Bnd terms} gives a consistent treatment of the boundary terms that are required by the closure of the action under the fermionic symmetry $\delta$.

\subsubsection{Absence of higher derivative corrections}

In this section we compute the renormalized action for the case when $c_I=0$, that is, when we do not have higher derivative corrections. 

Due to the form of the localization solutions it is convenient to introduce paracomplex variables defined as
\begin{eqnarray}
X^0_+=\Phi^{-1}e^{\alpha},\,X_-^{0}=\Phi^{-1}e^{-\alpha},\,|X^0|^2\equiv X^0_+X^0_-=\Phi^{-2}\nonumber\\
X^{I}_+=(\sigma+\chi)^Ie^{\alpha},\,X^{I}_-=(\sigma-\chi)^Ie^{-\alpha},\,|X^I|^2\equiv X^I_+X^I_-=\sigma_I^2-\chi_I^2\nonumber
\end{eqnarray}which are natural variables in theories with $SO(1,1)$ R-symmetry. With this parametrization the localization solutions are given by
\begin{eqnarray}
X^0_+=(*)+\frac{C^0}{\cosh(\eta)},\,X^0_-=(*)-\frac{C^0}{\cosh(\eta)}\nonumber\\
X^{I}_+=(*)+\frac{C^I}{\cosh(\eta)},\,X^{I}_-=(*)+\frac{C^I}{\cosh(\eta)}\nonumber
\end{eqnarray}Since no field has dependence on the fifth coordiante the \emph{Kaluza Klein reduction is exact}. The reduction goes much like in \cite{Banerjee:2011ts} except for the fact that the theory now has manifest $SO(1,1)$ R-symmetry.

 It is observed that the hypermultiplet lagrangian vanishes exactly on the localization locus, that is,
\begin{equation}
8\pi \mathcal{L}|_{\text{hyper}}=-\frac{1}{2}\Omega_{\alpha\beta}\varepsilon^{ij}\left\{ D_MA^{\alpha}_iD^MA^{\beta}_j-A^{\alpha}_iA^{\beta}_j\left(\frac{3}{16}R+2D+\frac{3}{4}T^2\right)\right\}|_{\text{loc}}=0
\end{equation} even though we were not able to localize in the hypermultiplet sector. We do not know if this is always true or just accidental. 

With the field redefinition
\begin{eqnarray}
&&X^{'0}_+=\frac{1}{2}iX^{0}_+,\,X^{'0}_-=-\frac{1}{2}iX^{0}_-\nonumber\\
&&X^{'I}_+=\frac{1}{2}X^{I}_+,\,X^{'I}_-=\frac{1}{2}X^{I}_-\nonumber\\
&&Y^{i0}_{\;\;j}=\Phi^{-2}\mathcal{V}^{\,\,i}_{j},\, Y^{iI}_{\;\;j}=2Y^{iI}_{\;\;j}|_{5d}+\chi\Phi Y^{i0}_{\;\;j}\nonumber\\
&&T=-8i\hat{T}\nonumber\\
&&A^I=A^{5dI}+i(X_+^{I}/X^{0}_+-X_-^{I}/X^0_-)(d\psi+B)\nonumber\\
&&A^0=B\nonumber
\end{eqnarray}
we can write the relevant non-zero part of the action  with a prepotential $F(X)$ given by
\begin{equation}
F^+(X)=\frac{1}{2}C_{IJK}\frac{X^{'I}_+X^{'J}_+X^{'K}_+}{X^{'0}_+}\nonumber
\end{equation}
as
\begin{eqnarray}
e^{-1} 8\pi^2\mathcal{L} =&& -i  \big(\partial_\mu X^{'A}_+ \,\partial^\mu  F^-_A -\partial_\mu X^{'A}_-  \,\partial^\mu F^+_A \big)\nonumber\\ 
&&+ \frac{1}{4}iF^+_{AB} (F^{-\,A}_{\mu\nu}-\frac{1}{4}X^{'A-}T^{-}_{\mu\nu})^2-\frac{1}{4}i F^-_{AB} (F^{+\,A}_{\mu\nu} -\frac{1}{4}X^{'A+}T^{+}_{\mu\nu})^2\nonumber\\ 
&&-\frac{i}{8}F^+_A(F^{+\,A}_{\mu\nu} -\frac{1}{4}X^{'A+}T^{+}_{\mu\nu})T^{+\mu\nu} +\frac{i}{8}F^-_A(F^{-\,A}_{\mu\nu} -\frac{1}{4}X^{'A-}T^{-}_{\mu\nu})T^{-\mu\nu}\nonumber\\
&& +\frac{1}{8}(-iF^+_{AB}+iF^{-}_{AB})\,Y_{ij}{}^A Y^{ij B}-\frac{i}{32}F^+(T^+_{\mu\nu})^2+\frac{i}{32}F^-(T^{-}_{\mu\nu})^2\nonumber\\
&&+\frac{1}{2}C_{IJK}\,d\left(t^IF^J\wedge A^K\wedge(d\psi+A^0)\right)\nonumber\\
&&+\frac{i}{8}C_{IJK}\,d\left(t^It^JF(A^0)\wedge A^K\wedge (d\psi+A^0)\right)
\end{eqnarray}where we have written the last two terms in differential form for easy reading, and defined $t=2\chi\Phi$. The total derivatives arise after expressing the Chern-Simons terms with four dimensional quantities.  Notice that the action above, apart from the total derivatives, has the same form as the one used for localization in \cite{Dabholkar:2010uh}. We borrow their results. 

In the absence of higher derivative corrections the boundary terms are given by
\begin{equation}\label{bnd terms RNaction}
S_{\text{bnd}}=\tilde{Q}\oint_{S_{\tau}} A+g\oint_{S_{\psi}}\sqrt{h(r_0)}A-\tilde{J}\oint_{S_{\tau}} B
\end{equation}with
\begin{eqnarray}
\tilde{Q}_I&=&\frac{1}{2}i C_{IJK}\sigma_*^J\sigma_*^K(5+\cosh(2\alpha^*)),\nonumber\\
g_I&=&i\frac{1}{4}C_{IJK}\sigma^J_*\sigma^K_*\sinh(2\alpha^*)\nonumber\\
\tilde{J}&=&\sinh(\alpha^*)\cosh(\alpha^*)^2C(\sigma^*)
\end{eqnarray}where once more we use $*$ to denote the on-shell value of the fields. The boundary action contributes not just to the on-shell renormalized action, that is, to the on-shell entropy, but also at the quantum level. The boundary quantum correction, which is linear in $C$, offsets an equal contribution coming from the bulk action. So overall, the renormalized action has, in a taylor expansion around the attractor background, no linear dependence in $C$,  which is equivalent to saying that the equations of motion are satisfied at $C=0$. In other words, 
\begin{eqnarray}
&&S_{\text{bnd}}|_{\text{Ren}}=6\pi C(\sigma^*)\cosh(\alpha^*)+4\pi C_{IJK}\sigma^I_*\sigma^J_*C^K\tanh(\alpha^*)\nonumber\\
&&S_{\text{total derivative}}|_{\text{Ren}}=-2\pi C(\sigma^*)\tanh(\alpha^*)^2\cosh(\alpha^*)-4\pi C_{IJK}\sigma^I_*\sigma^J_*C^K\tanh(\alpha^*)\nonumber
\end{eqnarray}Note that the quantum part of these two contributions cancel as expected. The constant piece on the other hand can be written as
\begin{eqnarray}
S_{\text{bnd}}|_{\text{Ren}}+S_{\text{total der}}|_{\text{Ren}}=&&6\pi C(\sigma^*)\cosh(\alpha^*)-2\pi C(\sigma^*)\tanh(\alpha^*)^2\cosh(\alpha^*)\nonumber\\
&&=-\pi q_Ie^I_{4d}+\pi Je^0_{4d}
\end{eqnarray}with $q_I,J$ the five dimensional electric and angular momentum charges respectively
\begin{eqnarray}
&&q_I=6C_{IJK}\sigma^J_*\sigma^K_*\nonumber\\
&&J=4C(\sigma^*)\sinh(\alpha^*)\nonumber
\end{eqnarray} and $e^I_{4d},e^0_{4d}$ the corresponding four dimensional electric fields
\begin{eqnarray}
&&e^I_{4d}=-\frac{\sigma^I_*}{\cosh(\alpha^*)}\nonumber\\
&&e^0_{4d}=\tanh(\alpha^*)\nonumber
\end{eqnarray}
 The bulk renormalized action on the other hand gives the contribution
\begin{eqnarray}\label{Sbulk renorm}
S_{\text{bulk}}|_{\text{Ren}}=&&-4\pi Q_I^{4d}C^I-4\pi iQ_0^{4d}C^0\nonumber\\
&&-2\pi i\left[F^+\left(X^{I*}_++C^I,iX^{0*}_++iC^0\right)-F^-\left(X^{I*}_-+C^I,-iX^{0*}_-+iC^0\right)\right]
\end{eqnarray}with
\begin{equation}
Q_I^{4d}=iF^-_A(X^{'*}_-)-iF^+_A(X^{'*}_+)\nonumber
\end{equation}the four dimensional charges, and $X^*$ denotes the on-shell values of the scalar fields. Note that even though we have written explicitly a term linear in $C$ in (\ref{Sbulk renorm}), the second term of the expression gives another with opposite sign, so overall we do not have linear $C$ dependence.

The four dimensional charges are computed to give
\begin{eqnarray}
&&Q_I^{4d}=-\frac{3}{2}C_{IJK}\sigma^J_*\sigma^K_*=-\frac{1}{4}q_I^{5d}\nonumber\\
&&Q_0^{4d}=iC(\sigma^*)\sinh(\alpha^*)=\frac{i}{4}J\nonumber
\end{eqnarray}which are related to the five dimensional charges by a proportionality factor.

Putting together boundary and bulk contributions we arrive at the final expression
\begin{equation}\label{entropy fnct}
S|_{\text{ren}}=\pi q^{5d}_I\phi^I+\pi J\phi^0-2\pi \left[F^+\left(\phi^I,1+\phi^0\right)+F^-\left(\phi^I,1-\phi^0\right)\right]
\end{equation}with 
\begin{equation}
\phi^I=-e^I_{4d}+C^I,\,\phi^0=e^0_{4d}+C^0.\nonumber
\end{equation}The index $I$ runs over the number of vector multiplets in the theory. On the other hand the renormalized action for four dimensional $\mathcal{N}=2$ theory, derived in \cite{Dabholkar:2010uh}, is
\begin{equation}\label{entr fnct 4d}
S|_{\text{ren }4d}=-\pi q_I\phi^I-4\pi i\text{Im}F\left(\frac{\phi^I+ip^I}{2}\right)
\end{equation}where $F(X)$ is the prepotential of the theory, and $q_I$, $p^I$ are the four dimensional electric and magnetic charges respectively. Here the index $I$ goes over the range $I=0\ldots n_V$. Note that in four dimensions we can turn on magnetic fluxes which appear in the renormalized action as the magnetic charges $p^I$. However in five dimensions  for an horizon with $S^3$ topology this cannot happen. In the case of the black ring the horizon has $S^1\times S^2$ topology which allows for dipole magnetic charges \cite{deWit:2009de}.

Under the analytic continuation $\phi^0\rightarrow i\phi^0$ and $J\rightarrow -iq_0$ the five dimensional renormalized action (\ref{entropy fnct}) acquires the form (\ref{entr fnct 4d}) for $p^0=1$ and $p^I=0$.

\subsubsection{On-shell renormalized action with higher derivative corrections}

The computation of the renormalized action in the presence of higher derivative terms is technically cumbersome and  for this reason it is still work in progress. It would be very interesting if we could put it in a form like (\ref{entropy fnct}), that is, as a function of the potentials $\phi$.
Notwithstanding this technical difficulty, we decided to present here the "tree-level" computation of the renormalized action. The interest is to show that this formalism agrees with the traditional Noether procedure, giving an entropy function "\`{a} la Sen", in the sense that the entropy equals the on-shell five dimensional lagrangian density.  

The final answer for the entropy after computing both the bulk renormalized action and boundary terms is
\begin{equation}
S=4\pi \cosh(\alpha)C(\sigma)-\pi c.\sigma \sinh(\alpha)^2\cosh(\alpha)
\end{equation}with $C(\sigma)=C_{IJK}\sigma^I\sigma^J\sigma^K$ and $c.\sigma=c_I\sigma^I$. This agrees with the result for the entropy computed using the Noether procedure (\ref{Euclidean charges}).

As an aside it is easy to show that the quantum contributions coming from each Wilson line cancel as 
\begin{eqnarray}
&&\text{Quantum contrb}=\lim_{\eta\rightarrow \infty}-i2\pi\hat{Q}_I\Delta(\chi \Phi)\tanh(\alpha)\cosh(\eta)+i4\pi\hat{g}_I\Delta(\chi\Phi)\sinh(\eta)\nonumber\\
&=&-\frac{3}{32}(2\pi)\cosh(\alpha)^2\tanh(\alpha)C+\frac{3}{32}(2\pi)\cosh(\alpha)^2\tanh(\alpha)C=0.
\end{eqnarray}This is in agreement with the fact that the renormalized action for the higher derivative terms does not contain terms linear in $C$, a fact observed in Mathematica. This confirms the validity of our boundary terms.

\section{Discussion and Conclusion}

In this work we considered the problem of computing the quantum entropy of five-dimensional rotating supersymmetric black holes using localization techniques. We focused on $\mathcal{N}=2$ supergravity, within the context of off-shell superconformal formalism, and showed using localization that, in the absence of higher derivative corrections, the quantum entropy function is the same as the four-dimensional counterpart after a suitable analytic continuation.

The inclusion of higher derivative corrections in the computation of the quantum entropy is more complicated. The reduction to four dimensions gives, besides the usual chiral content, corrections of the form Gauss-Bonnet in addition to D-type term corrections. Even though our analysis is independent of the higher derivative content, because it only relies on off-shell susy transformations, the computation of the renormalized action in a form that is dependent only on the unfixed modes revealed to be very difficult. 

In the case those corrections are absent we were able to compute the quantum renormalized action and showed that it matches with the four dimensional counterpart. However this is not  the full answer to the problem as there can be additional one-loop contributions. Within localization we used a partially fixed background together with some other gauge fixing conditions. As explained before it is not known or even if it is possible to construct an exact deformation in supergravity that we can use to localize the theory in a background independent way. Our method can only probe the perturbative part of this computation since it only requires the equations of motion that result from the localization action.  Instead we can think of an effective measure on the space of the localization solutions. To understand this we write the final answer as
\begin{equation}
d(Q,J)=\int \prod^I d\phi^I\mathcal{M}(\phi)e^{S_{ren}(Q,J,\phi)}
\end{equation}where $\mathcal{M}(\phi)$ stands for an effective measure on the space of $\phi$'s, the unfixed modes, and it should be computed from the one-loop effects we have just mentioned. Since we do not know how to compute the one-loop contribution from first principles we can try to determine the measure as in \cite{Dabholkar:2011ec}. The idea is to construct an induced metric on the space of collective coordinates using duality symmetry.

We know via the microscopic $4d/5d$ lift that the quantum entropies of four and five dimensional black holes are intimately related. For instance the microscopic BPS partition function of black holes in toroidally compactified four and five dimensional string theory are the same. By the equality of index and degeneracy for the near horizon degrees of freedom the black holes must have the same quantum entropy. We expect to explore this idea with concrete examples in a future publication.

The higher derivative content of the five dimensional theory can be used to address very interesting questions about the four dimensional black holes. Since supersymmetry is highly restrictive, not every four dimensional term can be uplifted to five dimensions. The converse is also interesting. The reduction to four dimensions gives rise to terms that cannot be written within the off-shell $\mathcal{N}=2$ formalism. For instance the reduced four dimensional action contains apart from the usual $4d$ $\mathcal{N}=2$ chiral higher derivative content, a Gauss-Bonnet contribution and D-terms. It was observed in \cite{deWit:2010za} that D-terms do not contribute to the on-shell entropy and later it was conjectured that their quantum contribution should also vanish \cite{Dabholkar:2011ec,deWit:2010za}. We believe that understanding how higher derivative terms contribute to the five dimensional quantum entropy we can shed light on the role of non-chiral corrections to the black hole entropy.

In this work we considered geometries that have an $AdS_2$ horizon. This is the near horizon geometry of a supersymmetric black hole. As discussed in the section \S\ref{on-shell solt} there is also an $AdS_3$ solution to the off-shell equations. Depending on how we identify the fifth coordinate we can have the near horizon geometry of a black ring or black string. The $AdS_3$ case is richer but at the same time more difficult. For instance we have to consider the contribution of $SL(2,\mathbb{Z})$ orbifolds of $AdS_3$, the usual BTZ black holes, to the path integral in a way consistent with localization. This has been attempted in \cite{Murthy:2011dk} but the answer is still unsatisfactory.

\subsection*{Acknowledgments}

It is a pleasure to thank  Atish Dabholkar and Sameer Murthy for useful  discussions and comments on the draft. I would also like to thank Mariano Chernicoff for help in revising the draft. The author would like to acknowledge the hospitality of CMAT (Centro de Matemática da Universidade do Minho), CFP (Centro de F\'{i}sica do Porto), and CERN where part of this work was completed. The research leading to these results has received funding from the European Research Council under the European Community's Seventh Framework Programme (FP7/2007-2013) / ERC grant agreement no. [247252].

\appendix
\section{Conventions}
\begin{itemize}
\item Minkowski metric has signature $(-,+,+,+,+)$.
\item In the lorentzian theory levi-civita tensors are defined as
\begin{equation}
\varepsilon^{abcde}=i\epsilon^{abcde}
\end{equation}with $\epsilon^{012345}=1$.
\item Antisymmetric tensors
\begin{equation}
\delta_{ab}\,^{cd}=\frac{1}{2}\delta_a^c\delta_b^d-\frac{1}{2}\delta_a^d\delta_b^c
\end{equation}

\item Cartan equations
\begin{equation}
de^a+\omega^{ab}\wedge e^b=0
\end{equation}
\item Curvature tensors
\begin{eqnarray}
&&\frac{1}{2}\mathcal{R}_{abcd}e^c\wedge e^d=d\omega^{ab}+\omega^{ak}\wedge\omega^{kb}\\
&&\mathcal{R}_{ab}=\mathcal{R}^k_{\,\,\,akb}\\
&&\mathcal{R}=\mathcal{R}^a_{\,\,a}
\end{eqnarray}
\item Weyl tensor
\begin{eqnarray}
&&R_{\mu\nu\rho\sigma}=\mathcal{R}_{\mu\nu\rho\sigma}-\frac{1}{3}(g_{\mu\rho}\mathcal{R}_{\nu\sigma}-g_{\nu\rho}\mathcal{R}_{\mu\sigma}-g_{\mu\sigma}\mathcal{R}_{\nu\rho}+g_{\nu\sigma}\mathcal{R}_{\mu\rho})+\frac{1}{12}(g_{\mu\rho}g_{\nu\sigma}-g_{\mu\sigma}g_{\nu\rho}\mathcal{R})\nonumber\\
{}\\
&&R_{\mu\nu\rho\sigma}R^{\mu\nu\rho\sigma}=\mathcal{R}_{\mu\nu\rho\sigma}\mathcal{R}^{\mu\nu\rho\sigma}-\frac{4}{3}\mathcal{R}_{\mu\nu}R^{\mu\nu}+\frac{1}{6}\mathcal{R}^2
\end{eqnarray}

\end{itemize} 

\section{Killing spinors}\label{KillingSpinors}
The Killing spinor equations are:
\begin{equation}
 \nabla_{\mu}\xi^i+\frac{i}{4}T_{ab}(3\gamma^{ab}\gamma_{\mu}-\gamma_{\mu}\gamma^{ab})\xi^{i}=0
\end{equation}with
\begin{equation}
 \nabla_{\mu}\xi=(\partial_{\mu}+\frac{1}{4}\omega^{ab}_{\mu}\gamma_{ab})\xi
\end{equation}
We choose the following representation for the $\gamma$ matrices:
\begin{equation}
 \gamma_0=\sigma_1\times \mathbb{I},\,\,\gamma_{1}=\sigma_{2}\times \mathbb{I},\,\,\gamma_{2}=\sigma_3\times \sigma_1,\,\,\gamma_3=\sigma_3\times \sigma_2,\,\,\gamma_4=-\gamma_0\gamma_1\gamma_2\gamma_3=\sigma_3\times \sigma_3
\end{equation}
\newline
For the metric
\begin{equation}
ds^2=\sinh(\eta)^2d\tau^2+d\eta^2+d\theta^2+\sin(\theta)^2d\varphi^2+\cosh(\alpha)^2\Big(d\psi+\cos(\theta)d\varphi-\tanh(\alpha)(\cosh(\eta)-1)d\tau\Big)^2
\end{equation}the Killing spinor equations are, in components:
\begin{eqnarray}
&&\partial_{\tau}\xi^i+\frac{1}{2}\cosh(\eta)\gamma_0\gamma_1\xi^i+\frac{1}{2}\sinh(\alpha)\sinh(\eta)\gamma_1\gamma_4\xi^i-\frac{1}{2}\cosh(\alpha)\sinh(\eta)\gamma^1\xi^i=0 \\
&&\partial_{\eta}\xi^i+\frac{1}{2}\sinh(\alpha)\gamma_{4}\gamma_{0}\xi^i+\frac{1}{2}\cosh(\alpha)\gamma_0\xi^i=0\\
&&\partial_{\varphi}\xi^i+\frac{1}{2}\cos(\theta)\gamma_2\gamma_3\xi^i+\frac{1}{2}\cosh(\alpha)\sin(\theta)\gamma_3\gamma_4\xi^i-\frac{1}{2}\sinh(\alpha)\sin(\theta)\gamma_3\xi^i=0\\
&&\partial_{\theta}\xi^i+\frac{1}{2}\cosh(\alpha)\gamma_4\gamma_2\xi^i+\frac{1}{2}\sinh(\alpha)\gamma_2\xi^i=0\\
&&\partial_{\psi}\xi^i=0
\end{eqnarray}

Note that the solution to these equations is related to the solution $\alpha=0$ by
\begin{equation}\label{R-symmetry}
 \xi^i=e^{\frac{1}{2}\alpha\gamma_4}\xi^i_0
\end{equation}
where $\xi^i_0$ solves the equations for $\alpha=0$.

\subsubsection{Solutions}

In the basis:

$ \\
\xi=\left(\begin{array}{c}
  a_1\\
a_2\\
a_3\\
a_4\end{array}\right)=a_1\left(\begin{array}{c}
               1\\
	      0\end{array}\right)\times \left(\begin{array}{c}
               1\\
	      0\end{array}\right)
           +a_2\left(\begin{array}{c}
               0\\
	      1\end{array}\right)\times \left(\begin{array}{c}
               1\\
	      0\end{array}\right) 
+a_3\left(\begin{array}{c}
               1\\
	      0\end{array}\right)\times \left(\begin{array}{c}
               0\\
	      1\end{array}\right)
+a_4\left(\begin{array}{c}
               0\\
	      1\end{array}\right)\times \left(\begin{array}{c}
              0\\
	      1\end{array}\right)\\
$
\newline

The Killing spinors are 
$\\
\\
 \xi^i_{++}=e^{\frac{i}{2}(\tau+\varphi)}
\left(\begin{array}{c}
e^{\alpha/2}\sinh(\eta/2)\sin(\theta/2)\\
e^{-\alpha/2}-\cosh(\eta/2)\sin(\theta/2)\\
e^{-\alpha/2}-\sinh(\eta/2)\cos(\theta/2)\\
e^{\alpha/2}\cosh(\eta/2)\cos(\theta/2)\end{array}\right)\;\;\;
\xi^i_{+-}=e^{\frac{i}{2}(\tau-\varphi)}
\left(\begin{array}{c}
e^{\alpha/2}\sinh(\eta/2)\cos(\theta/2)\\
e^{-\alpha/2}-\cosh(\eta/2)\cos(\theta/2)\\
e^{-\alpha/2}\sinh(\eta/2)\sin(\theta/2)\\
-e^{\alpha/2}\cosh(\eta/2)\sin(\theta/2)\end{array}\right)\;\;\;\\ \\
\\
\xi^i_{-+}=e^{-\frac{i}{2}(\tau-\varphi)}
\left(\begin{array}{c}
e^{\alpha/2}\cosh(\eta/2)\sin(\theta/2)\\
-e^{-\alpha/2}\sinh(\eta/2)\sin(\theta/2)\\
-e^{-\alpha/2}\cosh(\eta/2)\cos(\theta/2)\\
e^{\alpha/2}\sinh(\eta/2)\cos(\theta/2)\end{array}\right)\;\;\;
\xi^i_{--}=e^{-\frac{i}{2}(\tau-\varphi)}
\left(\begin{array}{c}
e^{\alpha/2}\cosh(\eta/2)\cos(\theta/2)\\
-e^{-\alpha/2}\sinh(\eta/2)\cos(\theta/2)\\
e^{-\alpha/2}\cosh(\eta/2)\sin(\theta/2)\\
-e^{\alpha/2}\sinh(\eta/2)\sin(\theta/2)\end{array}\right)
$

With this normalization equation (\ref{R-symmetry}) is satisfied.

\subsubsection{Some properties of the Killing spinors}
For the Killing spinor we are using to localize
\begin{equation}
\xi=\left(\begin{array}{c}
\xi^1\\
\xi^2
\end{array}\right)=
\left(\begin{array}{c}
\xi_{++}\\
\xi_{--}
\end{array}\right)
\end{equation}we have 
\begin{eqnarray}
&&V^a=\sum_{i=1,2}(\xi^i)^{\dagger}\gamma^{a}\xi^i=2(-\sinh(\eta),0,\sin(\theta),0,\cos(\theta)\cosh(\alpha)+\cosh(\eta)\sinh(\alpha))\label{KS2}\\ 
&&\braket{\xi}{\xi}=\sum_{i=1,2}(\xi^i)^{\dagger}\xi^i=2(\cosh(\alpha)\cosh(\eta)+\sinh(\alpha)\cos(\theta))\label{KS4}\\
&&|V|^2=\braket{\xi}{\xi}^2\\
&&\sum_{i=1,2}(\xi^i)^{\dagger}\gamma^{ab}\xi^i=0,\label{KS3}\\
&&(\xi^1)^{\dagger}\xi^2=(\xi^2)^{\dagger}\xi^1=(\xi^1)^{\dagger}\gamma^a\xi^2=(\xi^2)^{\dagger}\gamma^a\xi^1=0
\end{eqnarray}
We can also construct a triplet of real bilinears
\begin{eqnarray}
&&\Theta^1_{ab}=(\xi^2)^{\dagger}\gamma_{ab}\xi^1-(\xi^1)^{\dagger}\gamma_{ab}\xi^2;\nonumber\\
&&\Theta_{ab}^2=i\left((\xi^1)^{\dagger}\gamma_{ab}\xi^1-(\xi^2)^{\dagger}\gamma_{ab}\xi^2\right);\nonumber\\
&&\Theta^3_{ab}=i\left((\xi^2)^{\dagger}\gamma_{ab}\xi^1+(\xi^1)^{\dagger}\gamma_{ab}\xi^2\right);\nonumber
\end{eqnarray}They obey a "selfdual" equation in five dimensions
\begin{equation}
\Theta^i_{ab}=\frac{1}{2\braket{\xi}{\xi}}\epsilon_{abcde}\Theta^i_{cd}V^e;
\end{equation}
and are normalized as
\begin{eqnarray}
\Theta^i_{ab}\Theta^j_{ab}=4\delta^{ij}\braket{\xi}{\xi}^2.\nonumber
\end{eqnarray}They generate a complex structure in the sense that
\begin{equation}
\frac{1}{|V|^2}\Theta^i_{mn}\Theta^i_{np}=\delta_{mp}-\frac{V^mV^p}{|V|^2}
\end{equation}where the RHS is just the projector onto the space transverse to $V$.

\section{KK reduction and Susy variations}\label{app susy variations Weyl}

In this section we work out the susy variations for the Kaluza-Klein fields.
 
For the metric
\begin{equation}
ds^2=g_{\mu\nu}(x)dx^{\mu}dx^{\nu}+\Phi^2(d\psi+B)^2\nonumber
\end{equation}we compute the spin connections:
\begin{eqnarray}
 &&\omega^{4i}=\Phi^{-1}\big(\partial_{i}\Phi-\partial_{\psi}(\Phi B_i)\big)e^{4}+\partial_{\psi}(\Phi B_{[i})B_{j]}e^j+\frac{1}{2}\Phi F_{ij}(B)e^j\nonumber\\
&&\omega^{ij}=\bar{\omega}^{ij}-\partial_{\psi}(\Phi B_{[i})B_{j]}e^4-\frac{1}{2}\Phi F_{ij}(B)e^4\nonumber
\end{eqnarray}with $e^j$ and $\bar{\omega}^{ij}$ respectively the vielbein and spin connections of the four dimensional metric, and
$e^4=\Phi (d\psi+B)$. We have defined $F(B)_{\mu\nu}=B_{[\mu,\nu]}$.
We rewrite the spin connections as
\begin{eqnarray}
&&\omega^{4i}=\Phi^{-1}\mathcal{H}_i e^{4}+\frac{1}{2}\mathcal{F}_{ij}e^j\nonumber\\
&&\omega^{ij}=\bar{\omega}^{ij}-\frac{1}{2}\mathcal{ F}_{ij}e^4\nonumber
\end{eqnarray}with $\mathcal{H}_i=\partial_{i}\Phi-\partial_{\psi}(\Phi B_i)$ and $\mathcal{F}_{ij}=2\partial_{\psi}(\Phi B_{[i})B_{j]}+\Phi F_{ij}(B)$.

The susy variations of $\delta \psi^i_{\mu}$ and $\delta \psi^i$ in the Kaluza-Klein reduction (\ref{KK reduction}) become
\begin{eqnarray}\label{locEq:psi_mu}
&&\delta \psi^i_{\mu}=\partial_{\mu}\xi^i+3A_{\mu}\gamma^4\xi^i+\frac{1}{4}\bar{\omega}^{kl}\gamma_{kl}\xi^i+\frac{1}{4}\tilde{T}_{kl}\gamma^{kl}\tilde{\gamma}_{\mu}\xi^i+\frac{1}{2}\mathcal{V}_{\mu j}^{\;\;\;i}\xi^j\nonumber\\
&&-\frac{i}{2}\gamma_{\mu}\left(\eta^i+2i\gamma^4 \Slash A-\frac{i}{2}T_{kl}\gamma^{kl}\xi^i+\frac{i}{8}\mathcal{F}_{kl}\gamma^{kl}\gamma^4\xi^i\right)\\
\nonumber\\
&&\delta \psi^i=\frac{1}{2}\partial_{\psi}\alpha \gamma^4\xi^i-\frac{1}{4}\Phi \mathcal{F}_{kl}\gamma^{kl}\xi^i+\frac{1}{4}\Phi \tilde{T}_{kl}\gamma^{kl}\gamma^4\xi^i+\frac{1}{2}\gamma^4\gamma^i(\mathcal{H}_i+6\Phi\gamma^4A_i)\xi^i+\frac{1}{2}\mathcal{V}_j^{\;\;i}\xi^j\nonumber\\
&&-\frac{i}{2}\Phi\gamma^4\left(\eta^i+2i\gamma^4 \Slash A-\frac{i}{2}T_{kl}\gamma^{kl}\xi^i+\frac{i}{8}\mathcal{F}_{kl}\gamma^{kl}\gamma^4\xi^i\right)\label{locEq: psi}
\end{eqnarray}with
\begin{equation}
\tilde{T}_{ij}=3T_{ij}+\frac{1}{8}\epsilon_{ijkl}\mathcal{F}^{kl},
\end{equation}and $\tilde{\gamma}_{\mu}=e^i_{\mu}\gamma_i$. In addition we have 
\begin{equation}
\delta\check{\psi}^i=\frac{1}{2}\partial_{\psi}\alpha \gamma^4\xi^i
\end{equation}which vanishes for $\partial_{\psi}\alpha=0$ as expected.
We have put in evidence a common term in the susy transformations that we denote by $\tilde{\eta}$ 
\begin{equation}
\tilde{\eta}^i=\eta^i+2i\gamma^4 \Slash A\xi^i-\frac{i}{2}T_{kl}\gamma^{kl}\xi^i+\frac{i}{8}\mathcal{F}_{kl}\gamma^{kl}\gamma^4\xi^i.
\end{equation}The other susy transformations also contain a term proportional to $\tilde{\eta}$. If we choose appropriately $\eta$ we can have $\tilde{\eta}=0$. That is, we choose
\begin{equation}\label{eta}
\eta^i=-(2i\gamma^4 \Slash A\xi^i-\frac{i}{2}T_{kl}\gamma^{kl}\xi^i+\frac{i}{8}\mathcal{F}_{kl}\gamma^{kl}\gamma^4\xi^i)
\end{equation}Note that $\eta$ vanishes on-shell, this way respecting the boundary conditions.
\section{Localization equations}\label{App:locEqs}

In the following we develop some important expressions for the square of fermionic transformations. 
Consider the spinor $\chi$ with components
\begin{eqnarray}
 &&\chi^1=W\xi^1+K\xi^1+Y\xi^2+i B_a\gamma^a\xi^1+\frac{1}{2}\left(F_{ab}+i T_{ab}\right)\gamma^{ab}\xi^1\\
&&\chi^2=W\xi^2-K\xi^2+Z\xi^1+i B_a\gamma^a\xi^2+\frac{1}{2}\left(F_{ab}+i T_{ab}\right)\gamma^{ab}\xi^2
\end{eqnarray}
 with $K,Y,Z,B_a,C_{ab},T_{ab}$ real and $W$ complex. Then
\begin{eqnarray}
 \braket{\chi}{\chi}&=&\frac{\braket{\xi}{\xi}}{4}\left(F_{mn}+\frac{1}{2}\epsilon_{mnpqr}F^{pq}\frac{V^r}{\braket{\xi}{\xi}}+\frac{J}{\braket{\xi}{\xi}}\Theta^2_{mn}\right)^2+\frac{1}{2\braket{\xi}{\xi}}(V^mF_{mn})^2\nonumber\\
&&+\frac{\braket{\xi}{\xi}}{4}\left(T_{mn}+\frac{1}{2}\epsilon_{mnpqr}T^{pq}\frac{{V}^r}{\braket{\xi}{\xi}}+\frac{2}{\braket{\xi}{\xi}}B_{[m}V_{n]}+\frac{K}{\braket{\xi}{\xi}}\Theta^1_{mn}+\frac{L}{\braket{\xi}{\xi}}\Theta^2_{mn}\right)^2\nonumber\\
&&+\frac{\braket{\xi}{\xi}}{2}\left(\frac{V^m}{\braket{\xi}{\xi}}T_{mn}-B_n-\frac{1}{\braket{\xi}{\xi}}\text{Im}(W)V_{n}\right)^2\nonumber\\
&&+\frac{1}{2\braket{\xi}{\xi}}\big(B.V+\text{Im}(W)\braket{\xi}{\xi}\big)^2+\text{Re}(W)^2\braket{\xi}{\xi}\nonumber\\
{}
\end{eqnarray}where $V^a=\bra{\xi}\gamma^a\ket{\xi}$ and $V^2=\braket{\xi}{\xi}^2$. We have also denoted $Y=L+J$ and $Z=L-J$.

\subsection{Solving $\delta\psi_{\mu}=0$}

In this section we solve the gravitino equation studied in section \S\ref{Loceqs: non-rigid back}:
\begin{equation}\label{gravitino=0}
(\frac{1}{2}\partial_{\mu}\alpha+3A_{\mu})\gamma^4\xi(0)^i+\frac{1}{4}(\bar{\omega}^{kl}_{\mu}-\bar{\omega}^{0kl}_{\mu})\gamma_{kl}\xi(0)^i+\frac{1}{4}(\hat{T}_{kl}-\hat{T}^{0}_{kl})\gamma^{kl}\gamma_{\mu}\xi(0)^i=0
\end{equation}after setting the auxiliary fields to zero. In order to find a finite set of solutions we need to consider the analytic continuation of $\Delta\omega=\bar{\omega}^{kl}_{\mu}-\bar{\omega}^{0kl}_{\mu}$ to imaginary values. Interestingly this doesn't happen in the Minkowski case for which the authors of \cite{Gupta:2012cy} found $AdS_2\times S^2$ as the unique solution to the gravitino equation. We can study the equation in components or we can construct the auxiliary "lagrangian" $\delta\psi_{\mu}^{\dagger}\delta\psi_{\mu}$ whose vanishing locus is in one-to-one correspondence with the solutions we are looking for. In this sense we can use the formulas described previously for the square of susy transformations.

It is straightforward using the equations in \S\ref{App:locEqs} that
\begin{equation}
\Delta\omega_{ij}+\frac{1}{2}\epsilon_{ijpq5}{\Delta\omega}^{pq}\frac{V^5}{\braket{\xi}{\xi}}=0\nonumber
\end{equation}which are easily solved to give
\begin{equation}
\Delta\omega=0.\nonumber
\end{equation}From the identity
\begin{equation}
\gamma^{kl}\gamma^i=\gamma^k\delta^{li}-\gamma^l\delta^{ki}+\epsilon_{klij}\gamma^j\gamma^4\nonumber
\end{equation}we rewrite equation (\ref{gravitino=0}) as
\begin{equation}
C_{l}\gamma^4\xi(0)^i+B_{lk}\gamma^k\xi(0)^i+\frac{1}{2}D_{lk}\gamma^k\gamma^4\xi(0)^i=0
\end{equation}where we defined 
\begin{equation}
C_l=\frac{1}{2}\partial_{l}\alpha+3A_l,\,B_{lk}=\frac{1}{2}(\hat{T}_{kl}-\hat{T}^{0}_{kl}),\,D_{lk}=\frac{1}{2}\epsilon_{lkpq}(\hat{T}_{pq}-\hat{T}^{0}_{pq})\nonumber
\end{equation}

To solve this problem we look at equation
\begin{equation}
T_{mn}+\frac{1}{2}\epsilon_{mnpqr}T^{pq}\frac{{V}^r}{\braket{\xi}{\xi}}+\frac{2}{\braket{\xi}{\xi}}B_{[m}V_{n]}=0\nonumber
\end{equation}that comes from the localization action in the previous section. From the component $(m,n)=(i,j)$ we deduce
\begin{eqnarray}
&&-\frac{1}{2}\epsilon_{ijkl}D_{rk}v^l+B_{ri}v^j-B_{rj}v^i=0\\
&&\Leftrightarrow -\frac{1}{2}\epsilon_{ijkl}\epsilon_{rkpq}\Delta T_{pq}v^l+\Delta T_{ir}v^j-\Delta T_{jr}v^i=0
\end{eqnarray}with $\Delta T=\hat{T}-\hat{T}^{0}$ and $v=V/\braket{\xi}{\xi}$. It simplifies further to
\begin{equation}\label{Delta T eq}
\Delta T_{jl}v^l\delta_{ri}+\Delta T_{li}\delta_{rj} v^l+\Delta T_{ij}v_r+\Delta T_{ir}v_j-\Delta T_{jr}v_i=0
\end{equation}Contracting with $v^iv^r$ we find
\begin{equation}
\Delta T_{ij}v^i=0
\end{equation}Now if we contract equation (\ref{Delta T eq}) with $v^r$ and use the previous result we find
\begin{equation}
\Delta T_{ij}=0.
\end{equation}It follows immediately that 
\begin{equation}
\hat{T}=\hat{T}^{0},\, A_{\mu}=-\frac{1}{6}\partial_{\mu}\alpha
\end{equation}
\bibliographystyle{JHEP}
\bibliography{bigfinal}

\end{document}